\documentclass[]{aa}

\usepackage{graphicx}
\usepackage{txfonts}
\usepackage{booktabs}
\usepackage[colorlinks=true,allcolors=blue]{hyperref}

\usepackage{float}
\usepackage{natbib}
\usepackage{multirow}
\usepackage{multicol}
\usepackage{textcomp}
\usepackage{xcolor}
\usepackage{xspace}
\usepackage{verbatim}
\usepackage{booktabs}
\usepackage{microtype}
\usepackage[labelfont=bf]{caption}
\def\RJ{R_\mathrm{J}}

\def\Teff{T_{\text{eff}}}

\def\H2O{{H$_2$O}}
\def\co2{CO$_{2}$}
\def\ch4{CH$_{4}$}
\def\nh3{NH$_{3}$}
\def\ph3{PH$_{3}$}
\def\h2s{H$_{2}$S}
\def\prt{{\tt petitRADTRANS}\xspace}
\def\msim{{\tt MIRISIM}\xspace}
\def\Xmol{X_{\text{mol}}}

\DeclareUnicodeCharacter{2212}{-}
\begin{document} 

\titlerunning{Molecular Mapping with JWST/MIRI-MRS}
\authorrunning{P. Patapis et al}

\title{Direct emission spectroscopy of exoplanets with the medium resolution imaging spectrometer on board JWST MIRI}

\subtitle{I. Molecular mapping and sensitivity to instrumental effects}

\author{P.~Patapis \inst{1}
\and E.~Nasedkin \inst{2}
\and G.~Cugno\inst{1}
\and A.M.~Glauser\inst{1}
\and I.~Argyriou \inst{3}
\and N.~P.~Whiteford \inst{4,5}
\and P.~Molli\`{e}re\inst{2}
\and A.~Glasse \inst{6}
\and S.~P.~Quanz\inst{1}}

\institute{Institute for Particle Physics and Astrophysics, ETH Zurich, Wolfgang-Pauli-Str 27, 8093 Zurich, Switzerland
\\
\email{polychronis.patapis@phys.ethz.ch}
\and
Max-Planck-Institut für Astronomie, Königstuhl 17, 69117 Heidelberg, Germany
\and
Instituut voor Sterrenkunde, KU Leuven, Celestijnenlaan 200D, bus-2410, 3000 Leuven, Belgium
\and
Institute for Astronomy, University of Edinburgh, Royal Observatory, Blackford Hill, Edinburgh, EH9 3HJ, UK
\and
Centre for Exoplanet Science, University of Edinburgh, UK
\and 
UK Astronomy Technology Centre, Royal Observatory, Blackford Hill, Edinburgh EH9 3HJ, UK.
             }
  \date{Received June 29, 2021 ; accepted October 25, 2021 }
 \newcommand{\mum}{$\mu$m\xspace}
  \abstract
   {The Medium Resolution Spectrometer (MRS) of the Mid-Infrared Instrument (MIRI) on board the James Webb Space Telescope (JWST) will give access to mid-infrared (mid-IR) spectra (5-28 microns) while retaining spatial information. With the unparalleled sensitivity of JWST and the MIRI detectors, the MRS has the potential to revolutionise our understanding of giant exoplanet atmospheres.}
  {Molecular mapping is a promising detection and characterisation technique used to study the spectra of directly imaged exoplanets. We aim to examine the feasibility and application of this technique to MRS observations.}
   {We used the instrument simulator \msim  to create mock observations of resolved star and exoplanet systems. As an input for the simulator, we used stellar and planet parameters from literature, with the planet spectrum being modelled with the radiative transfer code \prt. After processing the raw data with the JWST pipeline, we high pass filter the data to account for the stellar point spread function (PSF), and used a forward modelling approach to detect the companions and constrain the chemical composition of their atmospheres through their molecular signatures. }
   {We identified limiting factors in spectroscopic characterisation of directly imaged exoplanets with the MRS and simulated observations of two representative systems, HR8799 and GJ504. In both systems, we could detect the presence of multiple molecules that were present in the input model of their atmospheres. We used two different approaches with single molecule forward models, used in literature, that are sensitive to detecting mainly \H2O, CO, \ch4, and \nh3, and a log-likelihood ratio test that uses full atmosphere forward models and is sensitive to a larger number of less dominant molecular species.}
   {We show that the MIRI MRS can be used to characterise widely separated giant exoplanets in the mid-IR using molecular mapping. Such observations would provide invaluable information for the chemical composition of the atmosphere, complementing other JWST observing modes, as well as ground-based observations.  }

   \keywords{   Methods: data analysis --
                Techniques: imaging spectroscopy --
                Planets and satellites: atmospheres --
                Planets and satellites: gaseous planets --
                Infrared: planetary systems --
                Space vehicles: instruments
               }

\maketitle

%

\section{Introduction}

The past decade has proved revolutionary for exoplanet science with thousands of exoplanets being detected spanning a large range of masses, temperatures, and separations. While detection surveys are still ongoing with missions such as the Transiting Exoplanet Survey Satellite \citep[TESS,][]{Ricker2015} and GAIA \citep{Gaia2016, Perryman2014}, detailed characterisation of exoplanets with spectroscopic observations allows one to probe the properties of their atmospheres,  as well as to have crucial links to the theory of exoplanet formation. 

The small subset of exoplanets discovered with direct imaging offers a very distinct sample for the population of giant gas planets. These objects are mostly very young and hot \citep{bowler_imaging_2016}, mainly due to observational bias; direct imaging is sensitive to the thermal emission of these still contracting planets, which are all detected in the near-infrared (near-IR). The young age of these planets makes them the best targets for testing formation theories amongst the currently observed exoplanet population \citep[e.g.][]{Spiegel&Burrows2012}. 
By comparing radiation of giant planets with evolutionary track predictions for hot- and cold-start models, one can aim to reconstruct the formation pathway that the planet underwent (e.g. core accretion, gravitational instability; \citealt{jr:Pollack1996, jr:Boss1997}) and its migration history \citep{2016_Mordasini}.

Furthermore, emission spectroscopy enables the characterisation of their atmosphere properties such as the pressure-temperature profile (P-T profile), molecular composition, and cloud properties \citep{madhusudhan_exoplanetary_2019}. This provides insight into the early atmosphere evolution of gas giants as they cool down over time \citep{bowler_imaging_2016}. By carefully constraining elemental abundance ratios such as carbon-to-oxygen ratio (C/O), nitrogen-to-oxygen ratio (N/O), and sulphur-to-oxygen ratio, it is possible to directly link current atmospheric properties to formation, as the abundance ratios of the proto-planetary disk at the location where the planet formed is encoded into its atmospheric chemical properties \citep{oberg_2011, Piso2016, Eistrup16, Turrini2021}. The history and evolution of forming planets as they interact with their natal environments can influence their chemical composition as well, by the amount of solids or gas that they accreted, or whether they migrated from their birth place to the observed location \citep{Turrini2021, Oberg2021}. For example one of the current theories behind the ammonia enrichment of Jupiter is based on formation beyond the N$_2$ ice-line \citep{Oberg2019}.

To date, all spectroscopic data of directly imaged exoplanets has been restricted to near-IR spectral bands \citep[0.95-3.3 \mum, e.g.][]{2013_Konopacky, Barman_2015_HR8799b, 2017_GRAVITY} from ground-based telescopes. With these measurements, the mass and radius cannot be empirically measured and are instead inferred from evolutionary and atmospheric models under certain assumptions such as hot and cold start \citep{bowler_imaging_2016}. Some exceptions based on the combination of direct imaging and stellar astrometry \citep[$\beta$ Pic b, HR8799][]{2018_Snellen_BetaPicb_mass, Brandt2021} and on stability requirement for the planetary orbits \citep[HR8799 system,][]{Wang_2018_HR8799}, offer important constraints for the models. The inference of other atmospheric properties such as temperature, surface gravity, clouds, and composition relies on the enhancing the spectral coverage and resolution, as well as improving atmospheric models that are fitted using Bayesian methods (e.g. \citealt{Samland_2017,Lavie_2017_HELIOS, Wang_2018_HR8799, Wang_2020_HR8799c, Nowak_2020_BetaPicb, Stolker2020, molliere_retrieving_2020, Wang_2021}). The lack of data for directly imaged exoplanets beyond 5 \mum \citep{petitDitDeLaRoche_midIRhr8799_2019, DelaRoche2021, Wagner2021} is mainly due to the dominating thermal background of earth's atmosphere. Mid-IR observations would provide valuable and complementary information to near-IR data to address model degeneracy in the effective temperature estimation, impact of clouds and sensitivity to different molecules \citep{madhusudhan_exoplanetary_2019, Bozza_molecules} and would give access to colder and older objects that are not detectable in the near-IR. The James Webb Space Telescope \citep[JWST,][]{Gardner_2006} will help revolutionise the characterisation of exoplanet atmospheres, followed soon by the next generation of extremely large ground-based telescopes (ELTs).      

The Mid-InfraRed Instrument \citep[MIRI,][]{Rieke2015a, Wright2015} on JWST is designed to serve many science cases, including exoplanet atmosphere characterisation. MIRI includes three observing modes, an imaging mode including coronagraphic filters, a medium resolution integral field spectrometer and a low resolution spectrometer. The Low Resolution Spectrometer \citep[LRS,][]{Kendrew2015} will mainly target transiting exoplanets with a resolving power of R$\sim$ 50-100 covering a wavelength range of 5-12 \mum. The MIRI Coronagraphic Imaging \citep{Boccaletti2015} offers three four quadrant phase mask (FQPM) coronagraphs at 10, 11, 15 \mum and a Lyot coronagraph at 23 \mum, for imaging gas giant exoplanets in these filters. Although the usual angular differential imaging \citep[ADI,][]{Marois2008} won't be as powerful as it has been for other facilities due to the small available telescope roll angle, the improved image stability and sensitivity compared to ground-based observations will provide the necessary contrast to observe most known objects, yielding 3 flux points from the FQPM coronagraphs, which can be used to constrain atmospheric models \citep{danielski_atmospheric_2018}. The Medium Resolution Spectrometer \citep[MRS,][]{Wells2015} that is the focus of this work, will be described in detailed in Section~\ref{sec:mrs_simulations}. It was not originally designed for exoplanet atmosphere characterisation, but rather for studying star formation and proto-planetary disks, as well as extra-galactic observations \citep{Rieke2015a}.

As the focus shifts to characterisation of exoplanets, there has been an increasing amount of interest by the exoplanet community in using integral field spectrometers (IFS) in the near-IR. Three of the current dedicated high contrast imaging instruments - SPHERE IFS \citep{SPHERE2019} at the Very Large Telescope (VLT), GPI \citep{GPIMacintosh} at the Gemini Observatory and ScEXAO CHARIS \citep{CHARIS2019_Currie} at the Subaru Telescope -  have implemented such modes, combining extreme adaptive optics and coronagraphy with a spectrometer. These instruments have only been able to achieve a spectral resolution of R $\sim$ 50-100, albeit at a high contrast. On the other hand, instruments such as VLT/SINFONI \citep[R$\sim$5000]{Eisenhauer2003}, Keck/OSIRIS \citep[R$\sim$4000]{Osiris2006} and VLT/MUSE \citep[][R$\sim$3000]{Bacon2010} have produced very interesting science output while lacking optimisation for direct imaging of planets. IFS instruments have allowed the detection of molecules in the atmospheres of HR8799 b/c \citep{2013_Konopacky, Barman_2015_HR8799b, PetitDitDeLaRoche2018}, $\beta$ Pictoris b \citep{Hoeijmakers2018, Rameau2021}, kappa And b \citep{todorov2016, Wilcomb2020} and HIP65426 b \citep{Petrus2020}, measurements of the radial velocity of HR8799\,b and c \citep{Ruffio2019} and the investigation of physical and chemical properties of the forming planet PDS70\,b and its surroundings \citep{Cugno2021}. This is made possible by taking advantage of the improved spectral resolution to disentangle the fundamentally different planetary and stellar signals, using cross-correlation techniques with forward models. In the near future, VLT/ERIS \citep{Davies2018} and Keck/KPIC \citep{Mawet2016} will provide upgraded spatially resolved spectroscopy on ground-based telescopes, and together with NIRSpec \citep{Bagnasco2007, ygouf2017_nirspec} on board JWST, will improve the characterisation of directly imaged exoplanets in the near-IR. Despite the large variety of existing and future near-IR instruments, the only planned mid-IR IFS are the MIRI-MRS slated to launch in 2021 and ELT/METIS \citep{METIS2021} that is expected to see first light in 2028.

In this work we investigate the potential of the MRS to detect molecules in the atmospheres of young giant exoplanets. In Section~\ref{sec:mrs_simulations} we introduce the instrumental specifications of the MRS relevant to observing faint signals and the tools to create simulated data of our targets. We then describe the cross-correlation data reduction steps for medium resolution IFS data in Section~\ref{CC_def}. Finally in Section~\ref{results} we demonstrate how the MRS can provide insights into the chemistry of directly imaged planet atmospheres, both through molecular mapping, and a likelihood based framework for the detection of various molecular species, and discuss the results in Section~\ref{discussion}.

\begin{figure*}[!ht]
    \centering
    \includegraphics[width=1.\hsize]{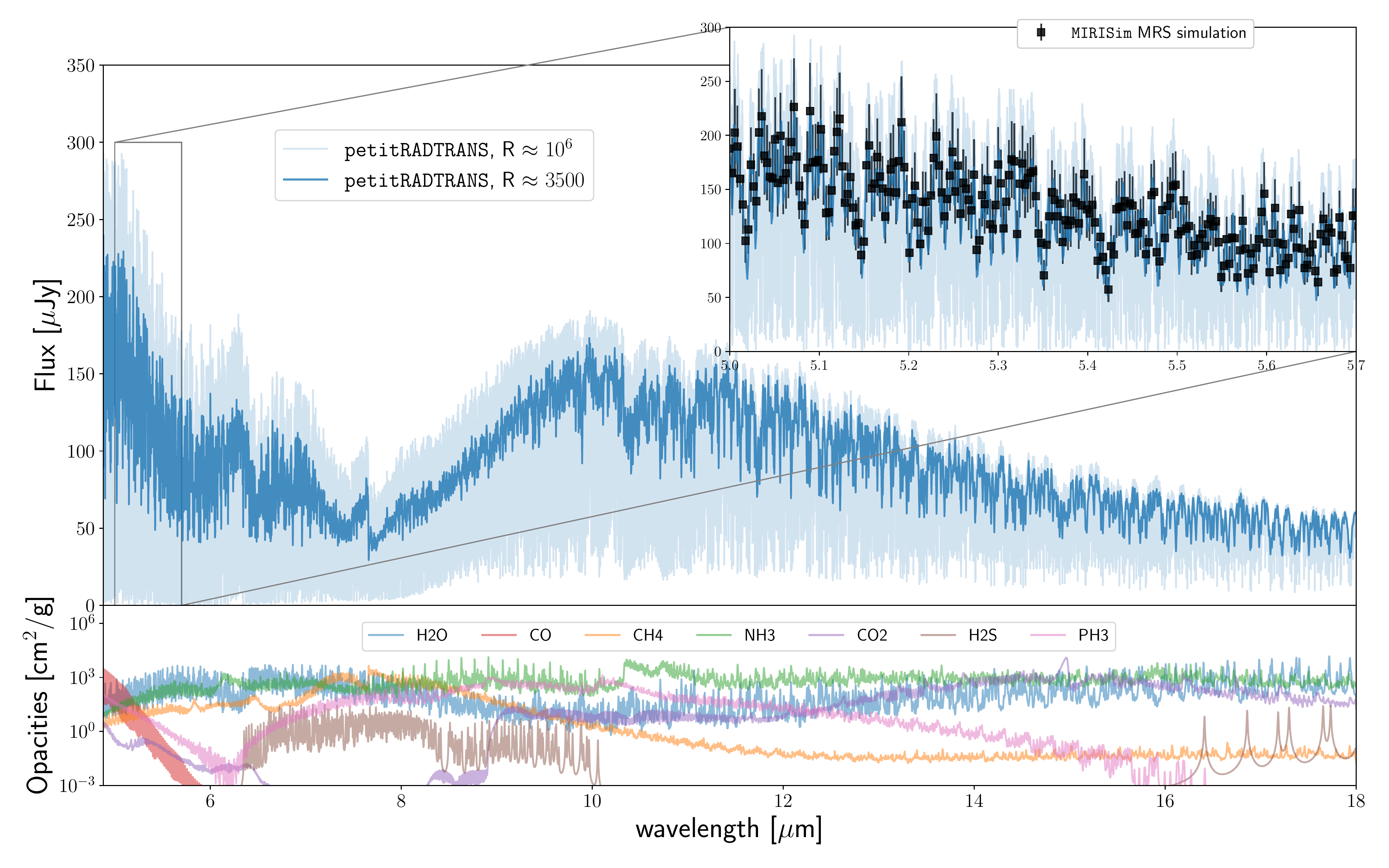}
    \caption{Top: Atmospheric model of HR8799 b calculated with \prt and simulated MRS data using \msim with realistic observation parameters, including all available noise sources. After the whole simulation chain and pipeline calibration steps we recover the input flux (see inset) by subtracting the stellar PSF with a simulation of HR8799 A without the planets. The error bars indicate the 3 $\sigma$ estimates from apertures with no planet signal at the same separation from the star as HR8799 b. Bottom: Opacity of different molecules as a function of wavelength in the mid-IR. Many molecules provide modulation in their opacity which will translate to detectable molecular features in their spectra.}
    \label{fig:MIRISIM example}
    \vspace{-1em}
\end{figure*} 
\section{Observing self-luminous companions with the MRS} \label{sec:mrs_simulations}

\subsection{MIRI medium resolution spectrometer}
The Medium Resolution Spectrometer \citep[MRS,][]{Wells2015} on MIRI  covers a wavelength range of 4.8 - 28.5 \mum with a resolving power of 4000 - 1500, decreasing with higher wavelength. It is divided into four overlapping wavelength channels (1-4) with two Si:As IBC detectors for channels 1/2 and 3/4 respectively, that are read out non-destructively \citep{Rieke2015b}. Additionally each channel is split into 3 spectral sub-bands (A, B, C) that all need to be observed in order to cover the full wavelength range. Since all channels can be exposed simultaneously the full wavelength coverage of the MRS requires just 3 observations, one for each sub-band. Being a IFS, the MRS retains both spatial coordinates with the end data product being a data cube where each pixel is a spectrum, denoted as a spaxel. The spatial resolution, field of view, and wavelength coverage of each sub-band can be seen in Table \ref{tab:specs}. These specifications provide some advantages over ground-based instruments of the same type, with a few key factors making the MRS unique. 

First, the low thermal background of JWST/MIRI will provide unprecedented sensitivity for faint sources in the $5-28$~\mum range \citep{Glasse2015}, compared to the limited sensitivity of ground-based exposures due to atmosphere's thermal background \citep[e.g. VLT/VISIR, VLT/NEAR,][]{petitDitDeLaRoche_midIRhr8799_2019, DelaRoche2021, Pathak2021}. Second, considering only the first 3 channels up to 18 \mum, since channel 4 suffers for a great drop in sensitivity, the continuous wavelength coverage is almost 10 times larger than current near-IR instruments, which will greatly benefit fitting atmosphere models and finding features of multiple molecular species. Lacking any mid-IR spectra, molecular features of giant planet atmospheres are currently not constrained. Thus, the MRS will add novel invaluable observational evidence to further improve and constrain existing atmospheric models. Additionally with the cross-correlation technique applied to IFS data we probe another aspect of the data, namely narrow molecular features from many different molecules \citep{Woitke2018}. Third, the contrast of exoplanets with respect to their host star is much more favourable in the mid-IR, with the stellar spectrum decreasing rapidly with wavelength ($\propto\lambda^{-4}$). At the same time, the features of typical stellar spectra at these wavelengths are originating from atomic absorption lines and will not correlate with the exoplanet atmosphere signature mostly dominated by molecules \citep{Bozza_molecules}. Finally, the point spread function (PSF) is expected to be much more stable in comparison to the near-IR, with none of the atmospheric turbulence seen in ground-based observing and well-characterised diffractive broadening dominating the spatial profile, compared to the alignment errors seen in the multi-element optical train which are more important at shorter wavelengths \citep{lightsey_jwst_wfe2014}. The gain in Strehl ratio in space will be an advantage for disentangling and modelling stellar and planet signal in our data.
 
Nonetheless, the complexity of the MRS requires precise calibration and pre-processing of the raw data. The Integral Field Units are based on a slicing optic (in contrast to a lens-let array implemented for instance in GPI and OSIRIS), which results in a different sampling for each spatial coordinate \citep{Glauser2010}. 
The optical design of the spectrometer introduces distortion of the dispersed light on the detector resulting in a non-orthogonal projection on it \citep{Wells2015}. 
The most prominent source of systematic error for the MRS is fringing caused by interference within the layers of the detector. It is expected to be more significant in the mid-IR as the wavelength of the light is comparable to the thickness of the detector layers and can cause up to 20\% flux variation as a function of wavelength \citep{Argyriou2020a}. Additionally, a straylight component \citep{Wells2015}, adds to the photometric uncertainty. Observations of point sources are always performed using dithered exposures in order to improve the spatial sampling, as  the spatial resolution is not Nyquist sampled for the short wavelength sub-bands \citep{Glauser2010}. Following multiple steps of the calibration pipeline \citep{Labiano2016} the data are calibrated (described in detail in Section ~\ref{jwst_pipeline}) and the detector images are reconstructed into a data cube, where an image is produced for each wavelength bin \citep{Wells2015, Labiano2021}. Residual systematic errors in each step of the calibration pipeline pose a significant obstacle to observations of very faint sources, such as exoplanets around a bright star, and careful correction of these residual effects might be critical to the success of the MRS as an exoplanet characterisation instrument. 

\begin{table}[h!]
         
\centering                          
\setlength\tabcolsep{2pt}
\caption[]{MRS specifications. FoV is the instrumental field of view. $\lambda$ denotes the spectral coverage and R refers to the average resolving power of the channel.}
\begin{tabular}{c c c c c}        
\textbf{Band} & \textbf{FoV ["]}$^a$ & \textbf{Spatial res ["]$^a$}  & \textbf{$\lambda$ [$\mu$m]$^b$} & \textbf{R}$^b$ \\ 
\toprule                        
1A  &  \multirow{3}{*}{3.2$\times$3.7} &  \multirow{3}{*}{0.196}
					  	  &4.88 - 5.77 &\\ 
1B   					& &.  &5.64 - 6.67 &3500\\  
1C   					& &	  &6.50 - 7.70 & \\    
   \hline                        
2A& \multirow{3}{*}{4.0$\times$4.8} &  \multirow{3}{*}{0.196}
					 			  &7.47 - 8.83 & \\ 
2B   					& &			  &8.63 - 10.19 &3200\\  
2C   					& &			  &9.96 - 11.77 & \\  
   								 
   	\hline                        
3A & \multirow{3}{*}{5.5$\times$6.2}  	&  \multirow{3}{*}{0.245}							
  					 		 &11.49 - 13.55 &\\ 
3B   					& &		  &13.28 - 15.66 &2500\\  
3C   					& &		  &15.34 - 18.09 &\\  
   	\hline                        
  4A & \multirow{3}{*}{6.9$\times$7.9}  &  \multirow{3}{*}{0.273}	
  					 		  &17.60 - 21.00 &\\ 
 4B  					& &		 &20.51 - 24.48 &1700\\  
 4C  					& &		 &23.92 - 28.55 &\\  
\bottomrule                                   
\end{tabular}
\tablebib{\tablefoottext{a}{\cite{Wells2015}.}\tablefoottext{b}{\cite{Labiano2021}.}}
   \label{tab:specs}
   \end{table}

\subsection{Observational limits} \label{sec:obs_limits}

The design and instrumental effects of the MRS place several constraints on the sample of giant exoplanets that can be characterised with it. In this section we present the main instrumental effects that dictate which planets are observable with the MRS: (i) bright source limit, (ii) spatial resolution, (iii) MRS sensitivity, and (iv) additional observing considerations related to the exoplanet system geometry and alignment with respect to the used dither pattern.  

(i) The MRS is expected to saturate on sources brighter than 4-5 Jy \citep{Glasse2015}. If the star is contained in the field of view of the scene that will be observed, the lack of a coronagraph means there is an upper limit to the length of integration that can be used without saturating the detector. In principle, the non-destructive readout of the detectors would allow for saturated pixels\footnote{For Cycle 1 observations saturation of the detectors is not permitted.}, with the pipeline using the partial ramp to fit the slope in post-processing. However, saturation for the MIRI detectors causes long lasting (up to several hours) persistence and is responsible for a series of effects that can affect the region around the saturated pixels, due to cross-talk between pixels \citep{Rieke2015b}. Reducing the length of the integration impacts the signal to noise ratio (S/N) of the planet and will result in longer exposure time (consisting of multiple integrations) being required.


(ii) The diffraction limited images delivered by the JWST have a diameter of the first dark Airy ring equal to 1.0 arcseconds at a wavelength of 10 \mum, making it  difficult to resolve planets with angular separations less than 0.5 arcseconds from their parent star. 

(iii) MIRI's sensitivity is unprecedented for the mid-IR.  However, we note that the MRS' high spectral resolution makes it less sensitive for detection of a planet's continuum spectrum, compared to the MIRI-LRS or imager, with the MRS limited to approximately 35 $\mu$Jy per spectral pixel (or equivalently a 18.2 K-band magnitude) for a 5$\sigma$ signal in 10,000 seconds \citep{Glasse2015}.  

(iv) There are two aspects of observing with the MRS that need special attention when observing spatially resolved planetary systems. First, the small Field of View (FoV - just 3.2 x 3.7 arcseconds in the short wavelength channel) makes the sky alignment of the instrument with respect to the exoplanetary system important. Especially when multiple companions are present, sub-optimal positioning of the field and choice of dither pattern might influence the signal of the planets. We built a visibility tool for the MRS that can help plan observations and choose a dither pattern illustrated in Fig.~\ref{fig:dither visibility} (Appendix~\ref{sec:visibility_tool}). Secondly, the detectors of MIRI are susceptible to persistence when bright sources are observed \citep{Rieke2015b}. Placing the planet at a position in the field where the star was located in a previous exposure will affect its appearance due to the additional latent image. Therefore the dither pattern choice should also consider this effect when planning observations.

From the Extrasolar Planets Encyclopaedia\footnote{\url{http://exoplanet.eu}} we collected all sub-stellar companions that fit the constraints outlined above. 
When limiting the sample to objects with reported mass in the planetary mass regime, the known objects that are observable with the MRS are shown in Table \ref{tab:targets} in  Appendix~\ref{ap:targets}. 
Two notable directly imaged exoplanet systems are missing: PDS70 b,c \citep{Keppler2018, Haffert2019} and 51 Eridani b \citep{Macintosh2015}. 
These are very interesting objects, the former being a forming planetary system whose companions are potentially surrounded by a circumplanetary disk \citep{Stolker2020, Cugno2021, Benisty21_pds70cpd}, and the latter one of the coolest imaged exoplanets with a temperature of $\sim$700 K \citep{Rajan_2017,Samland2017}. 
The PDS70 planets are too close to their host star to be spatially resolved, while 51Eri A is over the brightness limit of the MRS, saturating even with the minimum length of integration. Due to the non-destructive readout of the MRS detectors 51Eri b may become feasible when the technical problems of calibrating partially saturated images are well developed. 

\subsection{Two case studies for the MRS}
For the rest of this work we use two systems as a test-bed to illustrate the capability of the MRS to characterise exoplanets. 
These are the HR8799 system \citep{Marois2008, Marois2010} that hosts 4 known planets, and GJ504 b \citep{Kuzuhara2013, janson_direct_2013}. The two targets span a pivotal temperature regime that provides insights into the chemical processes of gas giant exoplanets, with one hosting hot planets and the other a cold planet.
The HR8799 planets have been extensively studied in the near-IR and are all estimated to have an effective temperature  between 1000-1200 K \citep{Marois2008, Bonnefoy_2016_HR8977, molliere_retrieving_2020}. One of the characteristics of this temperature regime for giant exoplanets is the change in the dominant carbon-bearing molecule in the observable photosphere, from CO to \ch4  for cooling temperatures \citep{madhusudhan_exoplanetary_2016}. While \cite{Barman_2015_HR8799b} detected CH$_4$ in HR8799\,b, more recently \cite{PetitDitDeLaRoche2018} and \cite{Ruffio_HR8799_2021} could not confirm this detection. Such open questions could be tackled with the extended wavelength coverage and sensitivity achieved by the MRS, placing limits on molecular abundances. 
The second system, GJ504 b has an effective temperature of $\sim$550K (150 K cooler than 51 Eri b), and was likely formed through gravitational instability \citep{bonnefoy_gj_2018}. 
Due to its low temperature it is expected to harbour both \ch4 (that has already been detected by \citealt{janson_direct_2013}) and \nh3 observable signatures in its SED, making GJ504 b a prime target to investigate the atmosphere physics of cooler gas planets. 

\begin{figure*}[!ht]
    \centering
    \includegraphics[width=1.\hsize]{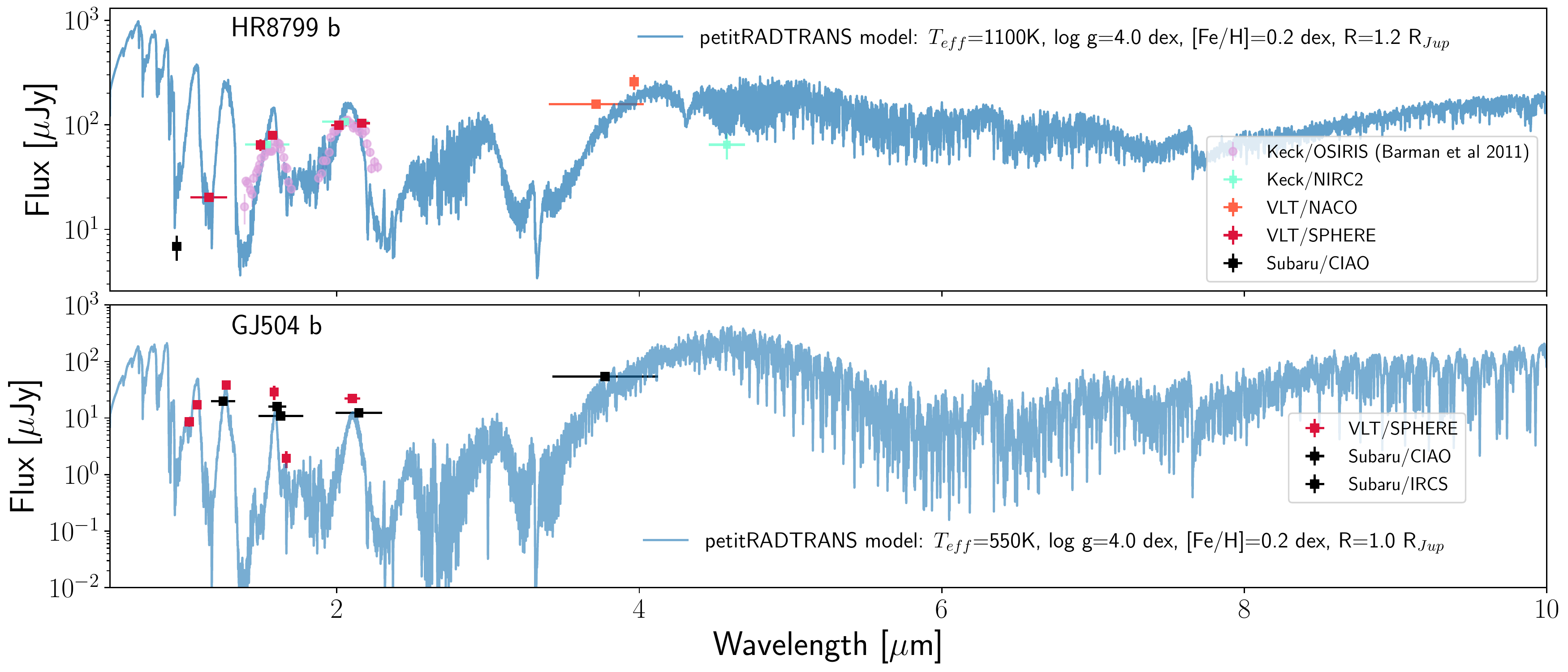}
    \caption{Atmospheric model calculated with \prt and archival near-IR data retrieved from \texttt{species} for HR8799 b (top) and GJ504 b (bottom). The models capture the observed spectral energy distribution and should provide a plausible emission spectrum of the planet atmosphere in the mid-IR.}
    \label{fig:pRT_data_comparison}
    \vspace{-1em}
\end{figure*}

\subsection{Emission spectra model}\label{sec:prt}
For calculating all emission spectra and forward models of the giant gas planets we use the 1D radiative transfer code \prt \citep{molliere_petitradtrans_2019, molliere_retrieving_2020}. 
With a user friendly and fast \texttt{python} implementation\footnote{\url{https://petitradtrans.readthedocs.io/en/latest/}}, this code takes as an input the thermal structure profile of the atmosphere (P$-$T profile), its gravity, and a list of continuum and line opacity sources together with the abundance of H, He, and the list of molecules included. It includes parameters for modelling clouds which we did not use in the main analysis. Clouds are expected to have distinguishable spectral features for mid-IR wavelengths at this spectral resolution \citep{Wakeford_silicateclouds_2015, Powell2019, Gao_Clouds_2020}, where broad features will be filtered out during the continuum removal, and higher frequency features would overlap other molecular features and could potentially be detected with cross-correlation. For the purpose of this work we consider clouds to overall reduce the contrast of molecular features, negatively impacting the detection of molecules.

One of the trade offs of this code is that the calculated emission spectra are not necessarily physical. In this work we assume a constant abundance of each molecule throughout the whole atmosphere altitude, for which the mass fraction value is arbitrarily set. To make sure that our models are at least plausible and representative of the atmospheres of known directly imaged planets we use the following inputs for the (i) p$-$T profile and the (ii) chemical composition of the atmosphere. 

(i) The p$-$T profile is taken from a grid of self-consistent models originally published in \cite{Samland_2017}. This grid was produced with \texttt{petitCODE} \citep{molliere_model_2015} and is parameterised by the effective temperature ($\Teff$), gravity ($\log(g)$), and metallicity (Fe/H). 

(ii) The authors of \prt offer a very useful tool for determining the equilibrium chemistry of an atmosphere, wrapped in a \texttt{python} package that is called \texttt{poor\_mans\_disequ\_chem}\footnote{\url{https://petitradtrans.readthedocs.io/en/latest/content/notebooks/poor_man.html}}. It is parametrised by effective temperature, gravity, C/O, and Fe/H, and even includes a simple quenching mechanism \citep{Molliere2017}. We use this tool in order to obtain a better physically grounded abundance value of each species in our models, calculating the abundance distribution of the different species as a function of altitude and selecting the abundance value for a given molecule at the pressure of $10^{-2}$ bar. We note that the abundances are still assumed to be vertically constant during our spectral modelling with the value found from the chemical equilibrium calculation above.

The list of parameters that are used in our \prt models are shown in Table ~\ref{tab:prt_params}. We kept the values of Fe/H and C/O fixed for all the planets in order to have more comparable atmospheres. To motivate that the calculated model spectra are a good description of the planets we simulate, we used the \texttt{species python} package \citep{Stolker2019} to retrieve near-IR data of HR8799 b and GJ504 b and plot them in Fig.~\ref{fig:pRT_data_comparison}. 

It should be noted that none of these models include the spin rate of the planet surface that would for higher resolution spectra affect the cross-correlation signal, seen as a broadening \citep{Snellen2014}. At the spectral resolution of the MRS this effect is negligible.

\begin{table}[h]
\caption{Atmosphere parameters used in our \prt simulation. }
    \label{tab:prt_params}
    \centering
    \begin{tabular}{l|c c c}
       \textbf{Parameter}  &  \textbf{HR8799 b$^a$} & \textbf{HR8799 c, d, e$^b$} & \textbf{GJ504 b$^c$}\\
       \toprule
      $R_{\mathrm{planet}}$ [$\RJ$] &    1.2 & 1.3             & 1.0\\
      Distance [pc]       &    41  & 41        &  17.5 \\
      log(g) [cgs]       &      4.0  & 4.0       & 3.9 \\
      $\Teff$ [K]    &  1100 & 1200 & 550 \\
      \hline
      Fe/H [dex]$^*$       & 0.2  & 0.2    & 0.2 \\
      C/O [dex]$^*$       & 0.4   & 0.4   & 0.4\\
      \hline
      \textbf{Molecule} & &\textbf{log($\Xmol$)} \\
      \H2O & -2.1 & -2.2 & -2.0 \\
      CO & -2.8 & -2.3 &  <-10 \\
      \ch4 & -2.4 & -2.7 & -2.3 \\
      \nh3 & -5.0 & -5.3 & -2.9 \\
      \h2s & -3.3 & -3.3 & -3.3 \\
      \ph3 & -5.0 & -5.1 & -8.0 \\
      \co2 & -5.3 & -5.0 & <-10\\
     \bottomrule
    \end{tabular}
    \tablebib{\tablefoottext{a}{\cite{Wang_2018_HR8799}.}\tablefoottext{b}{\cite{Wang_2020_HR8799c}.}\tablefoottext{c}{\cite{bonnefoy_gj_2018}.}}
    \tablefoot{\tablefoottext{*}{The parameters of Fe/H and C/O were kept constant in order to have a more comparable atmosphere chemistry.}
}
    
\end{table}

For the stellar models in our simulations we use the BT-NEXTGEN models \citep{Allard1997} accessed through \texttt{species} with the parameters of each host star taken from literature \citep{Marois2008, Kuzuhara2013}, shown in Table~\ref{tab:star_params}.

\begin{table}[h]
\caption{Parameters of stellar spectra.}
    \label{tab:star_params}
    \centering
    \begin{tabular}{l|c c }
       \textbf{Parameter}  &  \textbf{HR8799 A$^a$} & \textbf{GJ504 A$^b$}\\
       \toprule
      $R_\mathrm{star}$ [R$_\odot$] &    1.5 & 1.2 \\
      log(g) [cgs]       &      4.3  & 4.6  \\
      $\Teff$ [K]    &  7500 & 6300\\
     \bottomrule
    \end{tabular}
 \tablebib{\tablefoottext{a}{\cite{Gray1999}.}\tablefoottext{b}{\cite{Kuzuhara2013}}}   
\end{table}

\subsection{Data simulation - mock observations} 
We use the MIRI instrument SIMulator \citep[\msim,][version 2.3]{klaassen_mirisim_2020} to simulate realistic observations of the MRS. 
\msim is an end-to-end simulator that allows the user to define a custom astronomical scene and, given the observation parameters, it propagates the scene through a simulation where many known detector and instrument effects are modelled based on MIRI data obtained during several ground test campaigns.

For the simulations of our data we define a point source in the \msim scene for the star and each of the planets in the system. The location of the point sources is defined in RA and DEC with the centre of the MRS field as the zero-point. 
This is very convenient as the astrometry of exoplanets is often reported as $\delta$RA and $\delta$DEC with respect to the star, or as separation from the star and the position angle (PA) with respect to the north. Each point source is assigned a spectral energy distribution (SED) with an array of wavelengths (\mum) and fluxes ($\mu$Jy).

The observation parameters were optimised with the JWST Exposure Time Calculator\footnote{\url{https://jwst.etc.stsci.edu}}. 
The input parameters include detector readout mode (FAST or SLOW), number of groups per integration (NGROUP - effectively the number of readouts per integration), integrations per exposure (NINT), and dither pattern. 
For the MRS and our observations of exoplanets we always use FAST mode, with each frame exposed for 2.755 seconds. NGROUPS is optimised to maximise the number of groups such that the star does not saturate the detector. We choose NINT as necessary to achieve the desired S/N for the given exoplanet, while the adopted dither pattern is defined as described in Appendix~\ref{sec:visibility_tool}.  

We include all available instrumental and detector effects in our simulations, except cosmic rays, which have been shown to not behave as expected in \msim simulations and were disabled. 
The detector effects include photon noise, read noise, bad pixels, dark current, gain and detector flatness effect, non-linearity of the ramp, detector fringing, and a simple latency effect \citep{klaassen_mirisim_2020}. 
Instrumental effects include spatial and spectral distortion, PSF broadening in one of the spatial dimensions, and photon conversion efficiency, which includes the quantum efficiency of the detectors and the instrument optical chain transmission.

Using the procedure for creating MRS observations we simulated data for HR8799 and GJ504. For the HR8799 system we placed the star in the centre of the FoV and observed for 8 hours using NGROUPS=22, NINT=40, and the extended source 4 point dither pattern (Fig.~\ref{fig:dither visibility}). GJ504A was offset such that the star falls just out of the FoV in Channel 1 and GJ504 b is centred in the field. This was essential since the star is over the bright source limit of the MRS in Channel 1. We used NGROUPS=50, NINT=10, and the point source 4 point dither pattern for a total exposure time of 4.5 hours. In the inset of Fig.~\ref{fig:MIRISIM example} we show that the simulation correctly reproduces the input flux of HR8799 b. The flux was retrieved by simulating an observation with the same parameters as the HR8799 system but only containing the star, and then subtracting it to isolate the planet signal. Using an aperture with a radius encompassing 80\% of the PSF model we extracted the flux for each wavelength bin of the reconstructed cube. 

\subsection{JWST data reduction pipeline}\label{jwst_pipeline}

\msim delivers raw data in the form of FITS files exactly as the JWST data products that will be provided to observers on the Mikulski Archive for Space Telescopes (MAST). These detector raw data consist of one image for every frame, integration and exposure, in units of digital numbers (DN). DN represent the incident photons that were converted to electrons, read out by the detector, and stored in memory. In order to process these files and bring them to flux calibrated data cube form we use the JWST Data Reduction Pipeline\footnote{\url{https://jwst-pipeline.readthedocs.io/en/latest/}}. A detailed description of the pipeline steps applied to MRS raw data can be found in \cite{Labiano2016}.

In the first stage (\texttt{calwebb\_detector1}) the up-the-ramp samples of the non-destructive detector readouts in DN are transformed into count rate (DN/s). Here various detector level effects from the MIRI detectors are mitigated, such as: first and last frame effects, non-linearity of the ramps, detection of cosmic rays (not applicable in our case), and dark current correction. In the end a linear function is fitted to each pixel ramps and the slope is saved as the count rate of the given pixel, along with a data quality and error value.

Next, in the second stage of the calibration pipeline (\texttt{calwebb\_spec2}) all instrument specific physical corrections are applied. The goal is to produce a flux calibrated detector image for each exposure. These include the background subtraction, world coordinate system assignment, fringe correction, and spectro-photometric calibration.

Finally in the last stage (\texttt{calwebb\_spec3}) the different exposures for each dither position are reconstructed into a cube. This cube building is a critical step in the pipeline since it affects spatial and spectral resolution, introducing artefacts and correlated noise to the data \citep{Liu2019}. The current pipeline is using an exponentially decaying weight function to interpolate the flux of a given spaxel. The output of the JWST Data Reduction Pipeline consists of a data cube per MRS spectral sub-band.

\section{Cross correlation technique} \label{CC_def}

\begin{figure}[p] 
    \centering
    \includegraphics[width=\linewidth]{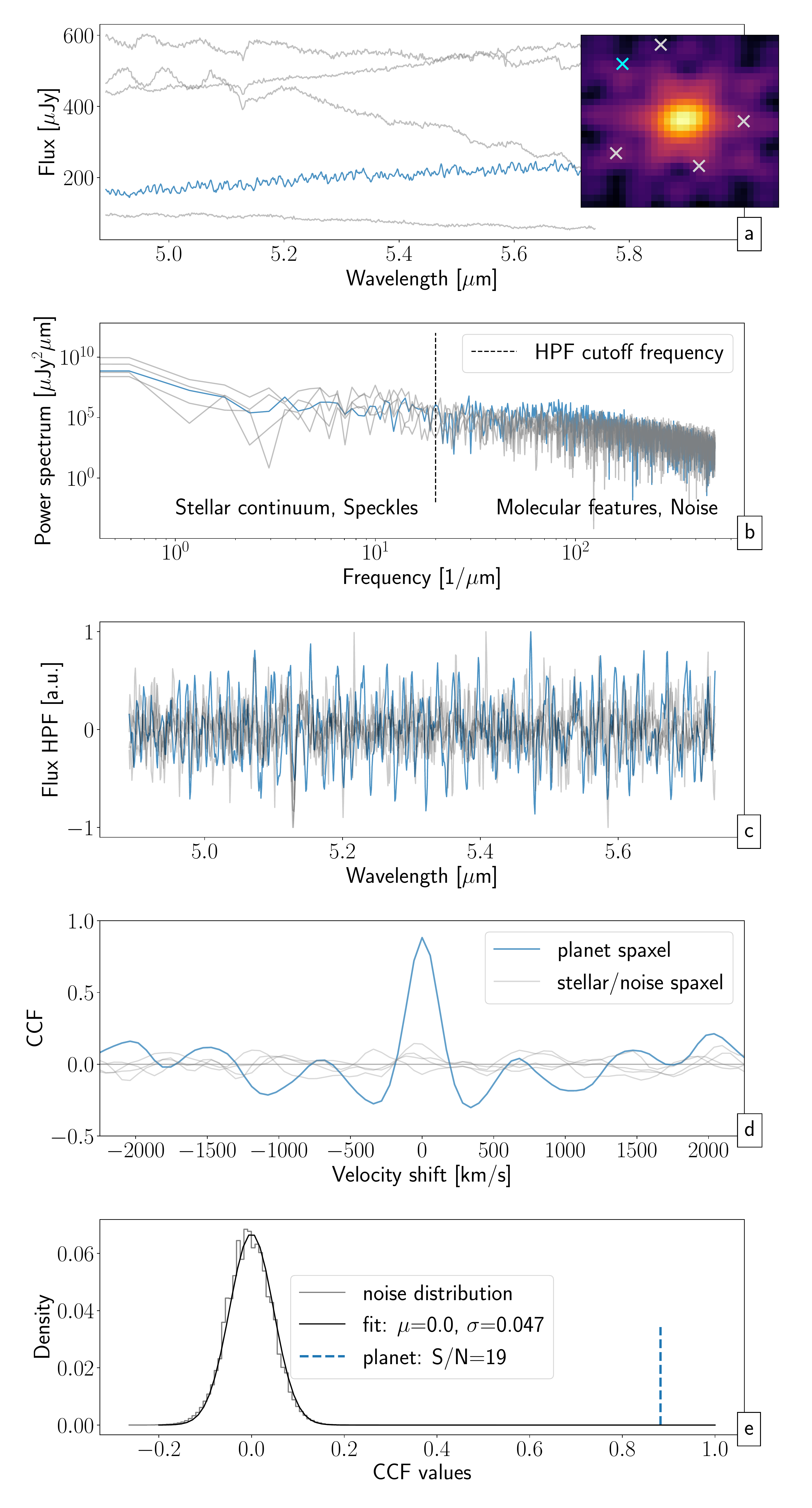} 
    \caption{Processing steps of MRS reconstructed cubes for the application of the cross-correlation technique. (a) Spectra at different locations in the FoV of sub-band 1A of the HR8799 observation were selected as shown in the inset (image in log scale). The spaxel containing planet signal (in blue) is not directly visible in the cube and is dominated by the stellar PSF. (b) Power spectrum of the selected spectra. By visualising the frequency domain we can see where the high pass filter operates and distinguishes continuum contribution dominated by the star and high frequency components that contain the molecular absorption lines and residual noise. (c) High pass filtered and normalised flux.  (d) Signal of the cross-correlation function where the input spectrum was used as a forward model. The planet spaxel clearly correlates with the forward model with a peak value of 0.88 (S/N $\sim$ 19). (e) S/N calculation for the cross-correlation method. All noise spaxels were considered and a Gaussian was fitted to the distribution to estimate the standard deviation that is used to estimate the S/N for a given cross-correlation value of the planet spaxel.}
    \label{fig:Highpass}
    \vspace{-1em}
\end{figure}

Cross-correlating a forward model of the spectrum with the IFU data has been very successful in isolating the exoplanet signal from the stellar PSF and the noise \citep{Hoeijmakers2018, Ruffio2019}. It is in essence a method of collapsing the data using weights given by a forward model, which results in boosting the S/N of spaxels where the spectra agree better with the forward model. This processing method is especially advantageous when high fidelity modelling, and subtraction, of the stellar PSF from the data down to the contrast level of the planet is not possible, as for example due to the lack of a coronagraph or instrument design. 

Applying cross-correlation to observations of exoplanets in the mid-IR with a space telescope has very clear advantages compared to similar observations from ground-based IFS. Indeed, (i) the lack of atmospheric speckles, (ii) the absence of telluric lines \citep{Hoeijmakers2018}, and (iii) the simplicity of the stellar spectrum in the Planck tail of the black-body with only a few atomic absorption lines will result in a low frequency component of each spaxel \citep{Ruffio2019}, even in the presence of low order wavefront errors from the telescope. 
Conversely, the exoplanet emission spectrum should contain many molecular absorption features and lines, with the opacity of different molecules being more prevalent at different wavelengths as shown in the bottom panel of Fig.~\ref{fig:MIRISIM example}. 
These features will translate to high frequency modulation of the spectra that can be separated from the dominating stellar signal in the data as we show below, for star-to-planet contrast ratio down to $\sim$10$^{-4}$ that is considered challenging even for coronagraphic observations with MIRI \citep{danielski_atmospheric_2018}.

For the data processing leading up to the cross-correlation we used a high-pass filter (HPF) by following a simplified version of \cite{Hoeijmakers2018}. The full processing used in this work consists of the following steps. The \texttt{JWST} pipeline was applied, producing a reconstructed cube with a given spatial and spectral sampling.
 For each spaxel the dominating stellar continuum was estimated by convolving the spaxel with a Gaussian and outliers were removed with sigma clipping of 10 standard deviations. This continuum was then subtracted from the spaxel, with only the high frequency component remaining in the data. The HPF spaxels were then normalised by dividing with their maximum absolute value. The forward model was convolved with a Gaussian to the average spectral resolution of the MRS for each sub-band and binned\footnote{The \texttt{python} package  \texttt{spectres} \citep{carnall2017spectres} was used to bin a high resolution spectrum to a new wavelength grid.} to the wavelength coordinates of the data, and then high-pass filtered and normalised in the same manner as the data. Finally, each spaxel was cross correlated\footnote{The \texttt{python} function \texttt{numpy.correlate} was used to perform the one dimensional correlation between two arrays.} with the forward model producing the cross-correlation function (CCF)
    $$c_{av}[k] = \frac{\sum_n a[n+k] \, v[n]}{\sqrt{a^T a\,v^T v}},$$ 
    where k is an index indicating the step of the CCF.

In Fig.~\ref{fig:Highpass} the processing is illustrated for a simulated data set where a random sample of spaxels was chosen (panel a) and their power spectral density (panel b), HPF spectrum (panel c) and cross-correlation with the forward model (panel d) is shown. The value for the Gaussian kernel standard deviation was chosen after varying the parameter and choosing the one producing the highest cross-correlation S/N (as defined in Section~\ref{sec:molecular_mapping}) when using the input model as the forward model. We expect this value to be a free parameter that would be optimised for each observation depending on the continuum of the stellar spectra and the PSF modulation as a function of wavelength. The spaxel containing the planet signal produces a peak in the cross-correlation while the rest of the spaxels indicate the noise floor from the various noise terms (Fig.~\ref{fig:Highpass}d). 

We note that the PSF of the star was not explicitly modelled but could be included in the forward model if there is an available stellar model, as for example shown in \cite{Ruffio2019}. Additional methods of dealing with remaining systematic residuals such as principal component analysis (e.g. \citealt{Cugno2021}) were not applied as the simulated data do not provide such effects, which are difficult to predict before having access to real on-orbit data.

After the last step in the data processing the final data product is a cube with the two spatial dimensions of the original cube, but with the spectral dimension now replaced by the values of the CCF for each velocity shift. The range of velocity shift for the CCF can be chosen but is in principle only limited by the wavelength range of the forward model.

\section{Analysis and results}\label{results}

\begin{figure*}[!ht]
    \centering
    \includegraphics[width=1.\hsize]{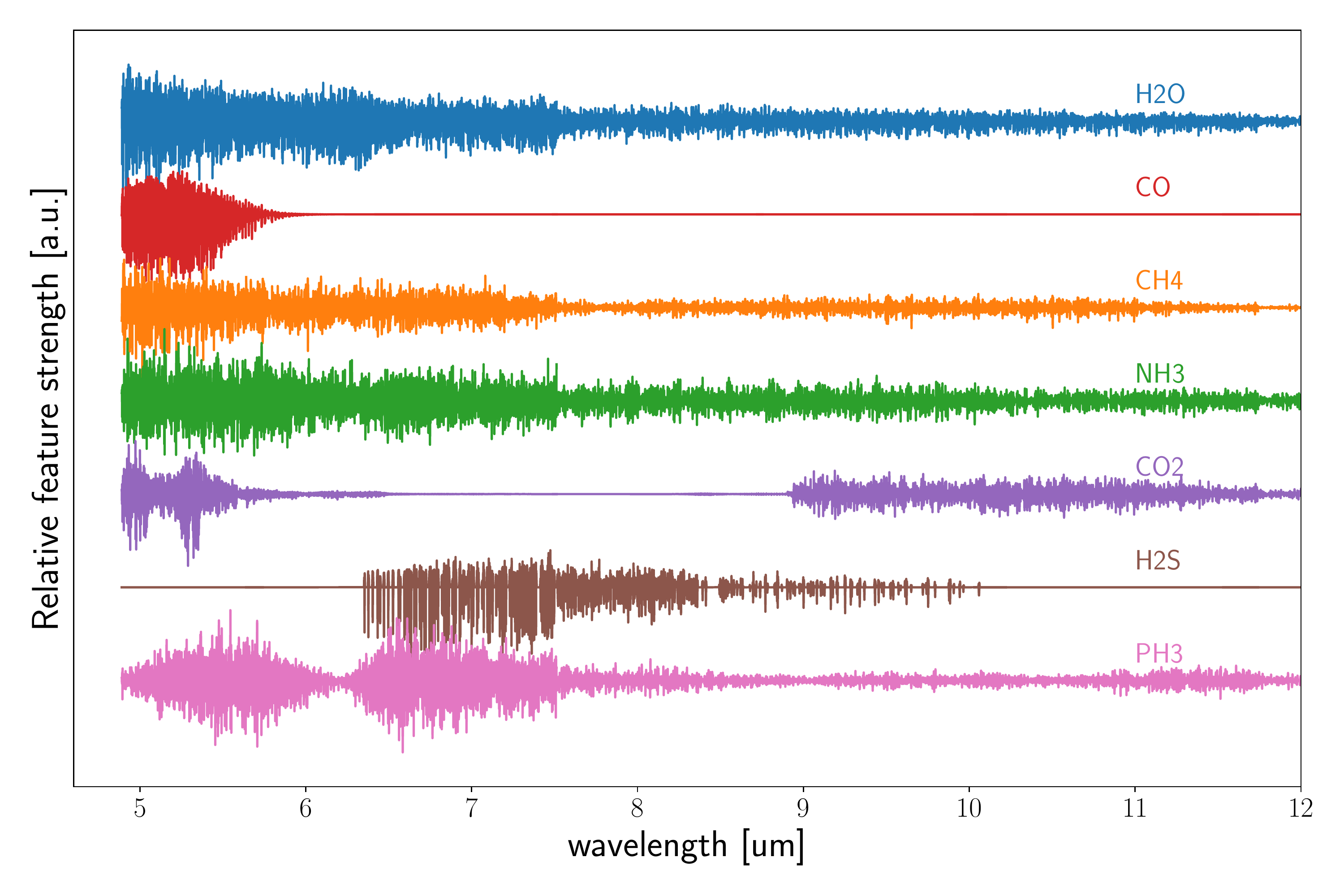} 
    \caption{Continuum subtracted molecule template spectra in the mid-IR at a spectral resolution of the MRS (R$\sim$ 3500 at 5 \mum) that were used in the analysis. The spectra are normalised to the absolute maximum value within the wavelength range of the MRS.}
    \label{fig:Molecule features}
    \vspace{-1em}
\end{figure*}

 \begin{figure*}[!ht]
    \centering
    \includegraphics[width=1.\hsize]{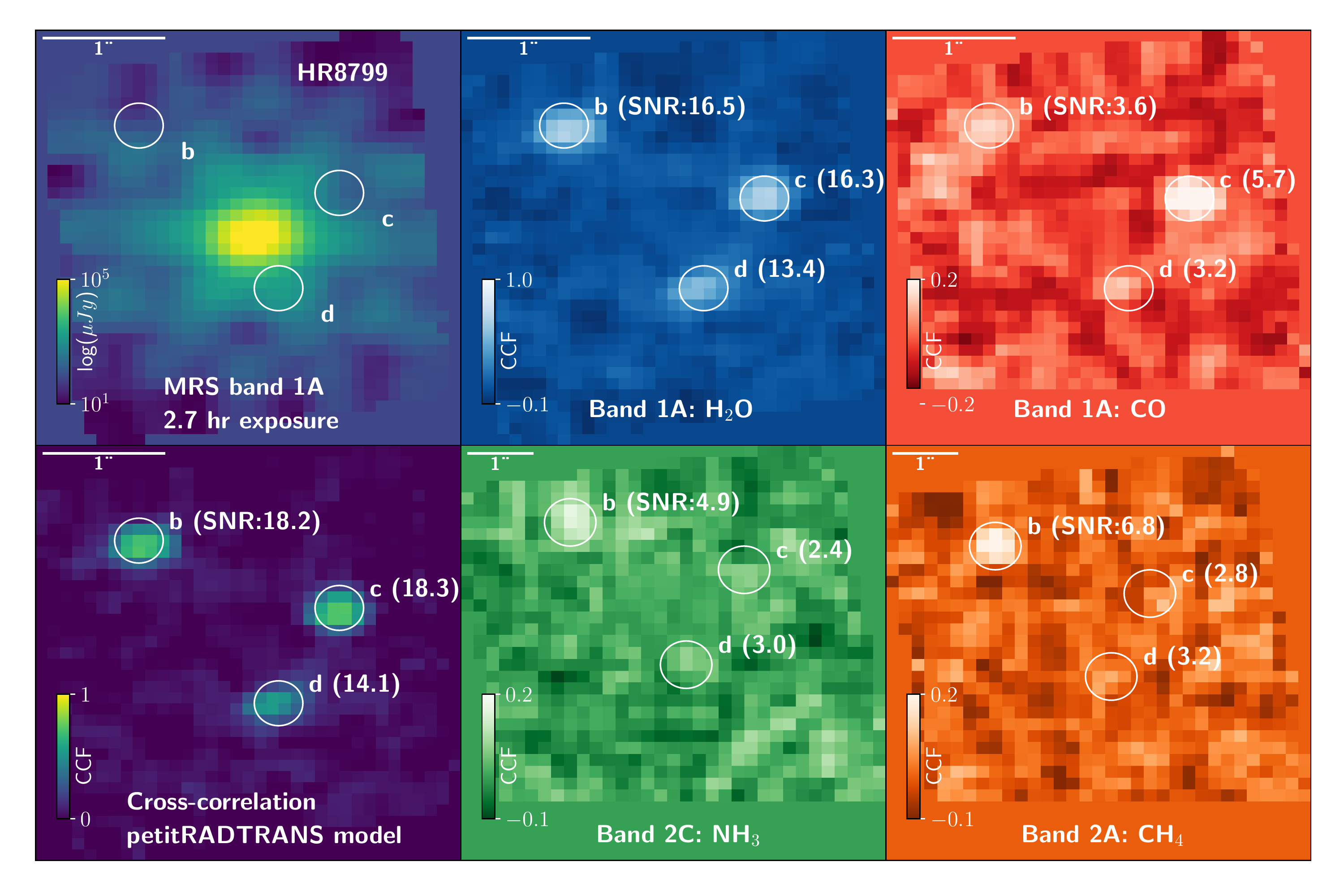} 
    \caption{Result of molecular mapping for a simulated observation of HR 8799 with the MRS for single sub-bands. \textit{First vertical column:} Single log-scaled frame from the flux calibrated and reconstructed cube, and the same frame cross-correlated with its input spectrum for sub-band 1A. \textit{Second and third vertical columns:} Cross correlation maps of filtered cubes with individual molecular templates correctly identify species of \H2O, CO for all planets, and \ch4, \nh3  for planet b. Values in parentheses indicate the S/N and the sub-bands with the highest S/N are plotted.}
    \label{fig:CC example}
    \vspace{-1em}
\end{figure*}

\begin{figure*}[!ht]
    \centering
    
    \includegraphics[width=1.\hsize]{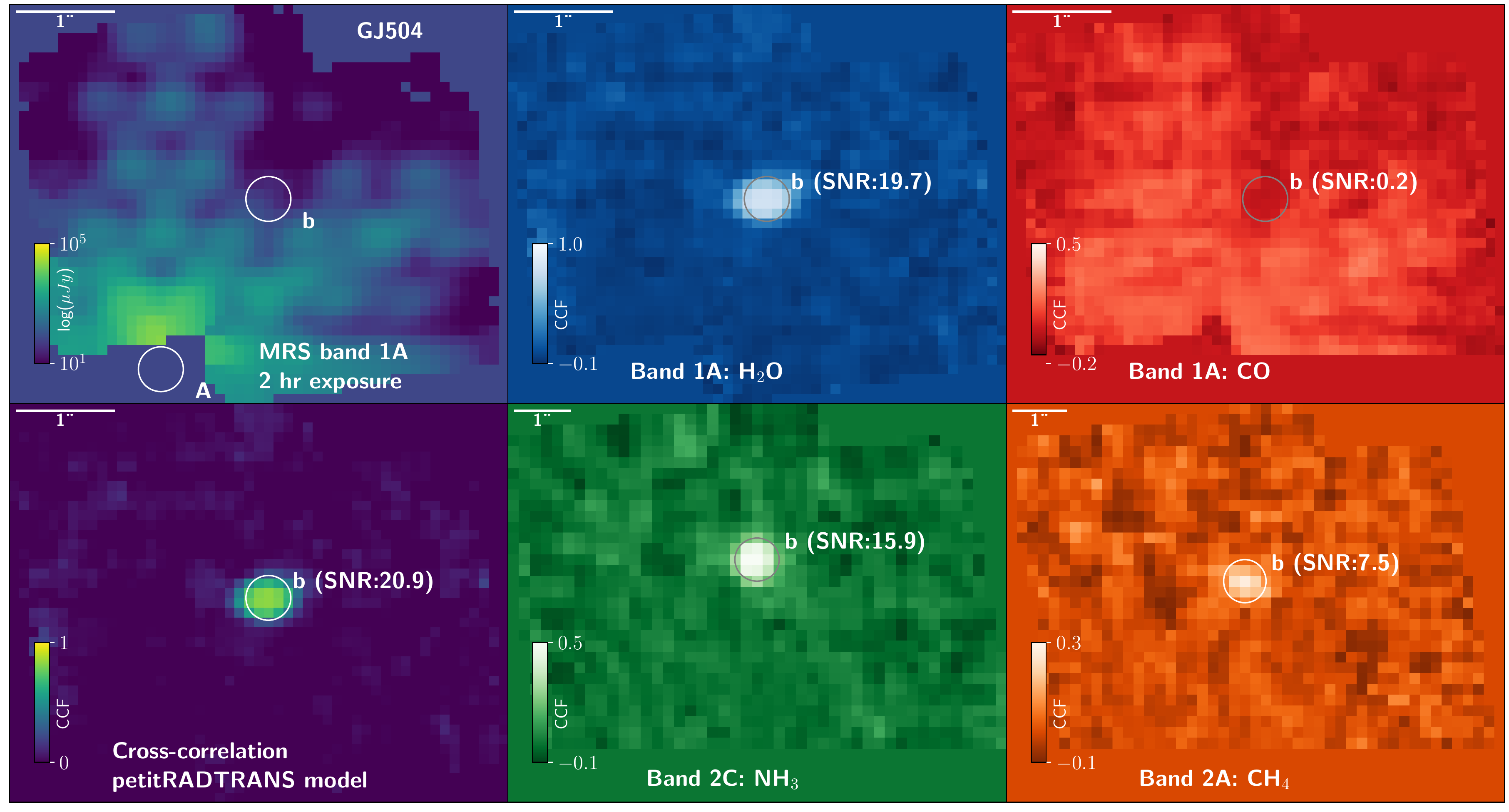} 
    \caption{Result of molecular mapping for a simulated observation of GJ504 b with the MRS for single sub-bands. \textit{First vertical column:} Single log-scaled frame from the flux calibrated and reconstructed cube, and the same frame cross-correlated with its input spectrum for sub-band 1A. \textit{Second and third vertical columns:} Cross correlation maps of filtered cubes with individual molecular templates correctly identify the abundant species of \H2O \ch4 and \nh3. No CO is detected (top right panel), as expected from chemical equilibrium models for cold objects as for instance GJ 504 b. Values in parentheses indicate the S/N and the sub-bands with the highest S/N are plotted.} 
    \label{fig:cc_gj504}
    \vspace{-1em}
\end{figure*}

\subsection{Molecular mapping in the mid-IR}\label{sec:molecular_mapping}

For the characterisation of these exoplanets with the MRS we started with the flux calibrated spectral cube and cross-correlated each spaxel with a template as described in Section~\ref{CC_def}. 
We used both the atmosphere model used as the input for the planet spectrum in the simulation, and molecular templates featuring the absorption signature of different dominating molecules such as \H2O, CO, \ch4, and \nh3. These molecule template spectra were again calculated with \prt by using the same P-T profile as for the input model but only considering the continuum contributors of H and He, together with the specific molecule opacity with a molecular mass fraction of $10^{-2}$. This spectrum may be nonphysical but it provides a template for the spectral features and lines that each molecule contributes to the emission spectrum of the exoplanet atmosphere. The continuum subtracted (as in Section~\ref{CC_def}) molecule templates with the P-T profile used for HR8799 b are shown in Fig.~\ref{fig:Molecule features}. 

 The S/N for the cross-correlation signal here is defined as in \cite{Petrus2020}. A t-test of the CCF signal at the radial velocity and position of the planet is performed with respect to the noise distribution of the CCF, illustrated in the Fig.~\ref{fig:Highpass}e. All pixels at a location not containing the planet signal and at radial velocity shifts larger than 500 km/s (equivalent to 10 pixels in the spectral direction) at the location of the planet were considered as noise samples. Before the calculation of the noise the auto-correlation of the forward model scaled to the signal of the CCF at the RV of the planet was removed was each spaxel, since it is evident that it dominates the noise for high correlation values, as seen in Fig.~\ref{fig:Highpass}d. The distribution of the noise sample is fitted with a Gaussian to estimate its mean and standard deviation. These are then used in the t-test to calculate the S/N as 
 \begin{equation}
 S/N = \frac{(\mu_{planet}-\mu_{noise})}{\sigma_{noise}},   
 \end{equation}
 
 where $\mu$ and $\sigma$ are the mean and the standard deviation, respectively. The planet CCF mean is calculated from an aperture centred on the planet.
 
The results are summarised in Fig.~\ref{fig:CC example} for HR8799 and Fig.~\ref{fig:cc_gj504} for GJ504. In Fig.~\ref{fig:CC example} the top left panel a single frame from the spectral cube is shown with the location of the planets annotated with circles.
Below, the cross-correlation with the forward model of the full atmosphere used as an input, reveals three of the four HR8799 planets: b, c, and d. The fourth planet HR8799 e is not detected since it is located too close to the star and its signal is lost under the noise limit. On the right, the four panels show the result of molecular mapping for the molecular species with the highest abundance. \H2O and CO  with a log mass fraction of -2.3 and -2.7 are detected with S/N of 16.5 and 3.6 for the HR8799 b planet. These molecules had previously been confirmed in \cite{Barman_2015_HR8799b} in the near-IR. \nh3 and \ch4 with log mass fraction of -5.0 and -2.4 are detected at an S/N of 4.9 and 6.8. Respectively for GJ504 b in Fig.~\ref{fig:cc_gj504}, molecules with high abundance such as \H2O, \nh3, and \ch4 are significantly detected with an S/N of 19.7, 15.9, and 7.5. These results provide detection of different species in the atmosphere of a planet without any prior knowledge of specific parameters for the planet atmosphere other than the line list that produced this molecule opacity in \prt. Although the P$-$T profile does affect the line contrast, the molecule template spectra are less sensitive on the specific choice we make due to the fact that we normalise it before cross-correlating with the data. It is important though to produce a template that highlights the features of the spectrum that the cross-correlation might pick up. This provides direct detection of molecules that are present in the atmosphere and can inform the next steps of the analysis. 

\subsection{Likelihood ratio test for molecule detection}

\begin{figure*}[t]
    \centering
    \includegraphics[width=\linewidth]{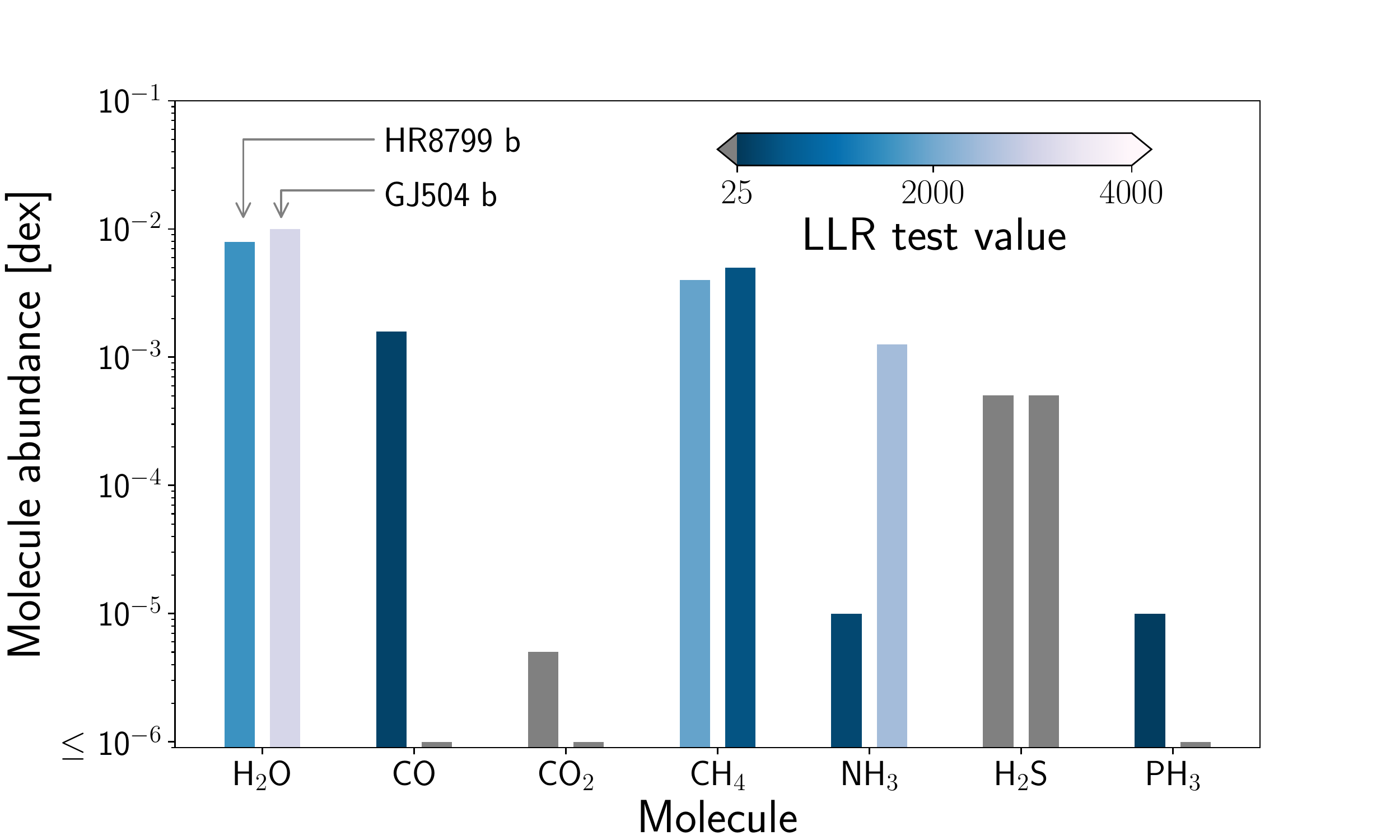}
    \caption{Log-Likelihood Ratio test statistic values for testing the existence of individual molecules in the atmosphere of GJ504 b and HR8799 b. The test statistic is calculated over all sub-bands of Channel 1 and 2. Values larger than 25 indicate a 5$\sigma$ detection for a $\chi^2$ distribution with 1 dof. The abundances of the individual molecules are taken from the input model shown in Table~\ref{tab:prt_params}.}
    \label{fig:llr_detections}
    \vspace{-1.5em}
\end{figure*}

While using a single molecule forward model to infer the presence of a molecule works for the dominant species, it might introduce a bias by not accounting for the impact of other species on the underlying spectrum, or not being sensitive enough to boost the planet signal of molecules with shallower features. As introduced by \cite{Ruffio2019} we used a likelihood ratio test for the detection of molecules, where we performed a model comparison between an atmospheric model including all species of the input model (Fig.~\ref{fig:Molecule features}) and one including all species except the molecule of interest. Hence we can test if adding a molecule to our atmosphere model significantly fits the data better than omitting it. Calculating a full atmosphere model implies that there are prior constraints for the effective temperature, P$-$T profile, and gravity of the planet. From self consistent atmospheric models, constraints from existing data and the result of the molecular mapping from Fig.~\ref{fig:CC example} one can obtain an initial estimation for the chemical composition of the atmosphere.

Using this forward model we located the signal in each cube and found its centroid by fitting a 2D Gaussian to the cross-correlation signal. For each sub-band we defined an aperture radius that encompasses 80\% encircled energy fraction of the PSF model, and extracted the mean flux within the HPF spaxels of the given aperture centred on the planet location.  We then transformed the cross-correlation value into a likelihood function as in \cite{2019_Brogi_Line} assuming a normal distribution for the noise: 
\begin{equation}
\log(\mathcal{L}) = -\frac{N}{2}\log\left(\sigma_{data}^2 - 2R(s) + \sigma_{model}^2 \right),
\end{equation}

where $\sigma_{data}^2$ and $\sigma_{model}^2$ are the variance of the data and the model respectively, and $R(s)$ is the cross-covariance coefficient for a given wavelength shift $s$. The cross-correlation is just the variance normalised cross-covariance. The difference of the log-likelihood of the two models that are being compared yields the log-likelihood ratio (LLR) statistic:
\begin{equation}
\Delta \mathcal{L} = -2\left(\log(\mathcal{L})_{full} - \log(\mathcal{L})_{-molecule}\right),
\end{equation}
where `full' indicates a forward model with all species and `-molecule' the full model without a given molecule. The resulting test statistic follows a $\chi^2$ distribution with 1 degree of freedom (dof), equal to the difference in free parameters of the two models \citep{2019_Brogi_Line}.  

We proceeded to perform the log-likelihood ratio test in order to test for the existence of molecules in the atmosphere. It is worth noting here that this method will work for the case of having detected the exoplanet in the cube with the forward model using molecular mapping. Moreover, it is only sensitive to atmospheres that contain molecular absorption features to detect since the cross-correlation is not sensitive to continuum modulation. This is especially the case for hotter giant exoplanets, where molecular features become less and less prevalent and are mainly limited to \H2O and CO, or for very cloudy objects, where thick clouds obscure molecular features \citep{madhusudhan_exoplanetary_2016}. 

For our two simulated data-sets we wanted to inspect which molecules would be detected in the ideal case where we use the input spectrum as the full model. This way we could probe the feasibility of detecting the different molecules, with the only limitation being set by the instrument's capabilities outlined in Section~\ref{sec:obs_limits}. Omitting just one molecule and keeping the rest of the molecules constant could still fit better and introduce a bias when applying this to real data, where we do not know the true values of the rest of the parameters. Therefore the LLR statistic values represent an upper limit of the detection sensitivity of the instrument to a specific molecule that is present in the atmosphere.

In Fig.~\ref{fig:llr_detections} we present the value of the log-likelihood ratio test summing over all sub-bands in Channel 1 and 2 (4.8 - 12 \mum) for a list of molecules that could be expected in such an atmosphere. Values, larger than 25 (critical value such that cumulative distribution of $\chi^2$ with 1 dof is 0.999997, equivalent to the 5$\sigma$ detection threshold), of the test statistic can be interpreted as molecules with significant contribution to the atmosphere spectrum, conditional on the full forward model that is used as a reference. Values around zero indicate that no significant change in the log-likelihood of the cross-correlation is observed, being the case for molecules that either do not have any features at these wavelengths (e.g. CO, only has a feature in sub-band 1A), or weak spectral features that are suppressed due to other more dominant, broadband absorption features present at the same wavelength. This is seen for example of \ch4 that is almost equally abundant in the two planets but detected at very different strengths due to the presence of much more \nh3 for the colder GJ504 b. Interestingly, \ph3 is detected for HR8799 b despite its low abundance of $10^{-5}$. The detection takes place in sub-bands 2B and mainly 2C, and is attributed to the fact that there are many features of \ph3 at these wavelengths that are not suppressed by other molecules with the given composition. This contrasts the non detection of \h2s which is much more abundant, but its weak spectral signature coincides with \ch4 and \nh3. Finally, negative numbers of the LLR would indicate that the model without the molecule significantly fits the data better. This is observed in our results in sub-band where multiple dominating absorbers are present simultaneously.  The LLR values of each individual sub-band are listed in Tables \ref{lLR_hr8799b} and ~\ref{lLR_gj504b}. This approach will work very well with the MRS since it changes the detection of low S/N individual molecular templates to evaluating and comparing models of high S/N of the cross-correlation with the full atmospheric model over the wide wavelength coverage offered by the instrument.

\section{Discussion}\label{discussion}

\subsection{Molecular mapping with the MRS}

As the only mid-IR IFS until ELT/METIS, the MIRI-MRS is uniquely positioned to contribute to the characterisation of giant exoplanet atmospheres. We show that with the cross-correlation technique, multiple molecular species are detected simultaneously at high confidence for two known systems, taking advantage of the broad wavelength coverage and moderate spectral resolution. The detection sensitivity of trace species greatly depends on the prior knowledge of the system and will vary as a function of the composition and relative abundances of different species. The mid-IR provides a unique discovery space for constraining the chemical composition of exoplanets. We expect \nh3 being one of the prevalent molecules in the mid-IR, which is supported by our results, being detected in both GJ504 b and HR8799 b. Unexpectedly, \ph3 is also detected in sub-bands 2B and 2C for HR8799 b using the likelihood ratio test, despite its weak molecular signature. Based on chemical equilibrium models, this molecule could be expected in such an atmosphere \citep{Wang2017_ph3} and if proven detectable it would be the first detection of \ph3 in an exoplanet. For planets such as HR8799 b, \ph3 would be well mixed and could trace vertical mixing in the atmosphere \citep{Fletcher2009}, as well as tracing the total phosphorus abundance of the atmosphere \citep{Visscher_2006}.

We explored the S/N as a function of the total exposure time of the full model and of different single molecules by using the output of the MRS simulation, only keeping part of the total number of integrations in increasing steps, and re-running the JWST pipeline each time. Because of the dependence of detection on the number of available features and the composition of the atmosphere, the time required to obtain a significant detection with molecular mapping can be very different for each molecule, even for molecules of comparable abundance in the atmosphere. Molecules with many, deep features such as \H2O and \nh3 are already detected from the first few integrations (in a timescale of a few tens of minutes, depending on the contrast of the planet). Molecules such as \ch4 and \ph3 can be affected more, both by the abundance of other species and the number of groups/integration possible due to the bright source limit, requiring multiple hours of observations to constrain them well.

Given the simulations presented in this work, we observe that close to the star the stellar photon noise is dominating the cross-correlation signal, while further out the detector noise and photon noise of the given planet are the limiting factors for the cross-correlation S/N. For all separations the spectral resolution of the instrument and the atmospheric composition of the planet in question will greatly impact the S/N, since the cross-correlation signal is dependent on both quantities. To test this we simulated an observation of just the HR8799 planets without the star and measured the S/N. As expected, the S/N was only slightly improved for the outer planet b, while for HR8799 c, d, and e, which are increasingly closer to the star the change in S/N was significant. Removing the star from the simulations allowed for the detection of the HR8799 e planet, that is in reality not considered feasible, and doubled the S/N of the HR8799 d planet. The analysis does not account for systematic errors that the in-orbit state of the MIRI instrument might introduce when observing faint sources or subtle effects the simulations might miss. For example, the inability to obtain deep integrations due to saturation of the detector and a whole series of related effects, such as persistence, that come into play when observing bright stars, might prove a fundamental limit to which targets are observable in an acceptable amount of telescope time.

Regarding the proposed log-likelihood ratio test for the detection of molecular signatures, introducing multiple species complicates the forward modelling and adds a few caveats that should be considered. Assuming a given molecule is actually present in the atmosphere and the full forward model including this molecule produces a high correlation, it can still occur that the model without the molecule produces a higher log-likelihood, hinting towards the molecule not being significant. This could be explained either by the dependence of molecular features on the P-T profile or the presence of another dominating absorber that was not accounted correctly in the initial model. The P-T profile has a large impact on the strength of the molecular feature line-to-continuum ratio and a profile that deviates significantly from the truth could distort the result of the log-likelihood ratio test. In the latter case, if there are two molecules that overlap in a given wavelength range, the removal of one from the model could enhance the features of the other leading to a better fit and an overall higher value of the LLR test statistic. Given these caveats, we should consider the results of the LLR result as an upper limit for the detection of molecules in the atmospheres of the simulated planets.

One of the next frontiers in characterising atmospheres of self-luminous exoplanets in the next years is the modelling of clouds. Whilst clouds have been studied in more detail for brown dwarfs \citep[see references covered in ][]{molliere_retrieving_2020}, comparing models to exoplanet spectra has been more challenging, but with increasing amounts of effort dedicated towards this problem \citep{Rajan_2017, Samland_2017, Charnay_2018, molliere_retrieving_2020, Burningham2021}. To test how clouds might affect our results we assumed a cloud model as retrieved by \cite{molliere_retrieving_2020} for the HR8799\,e planet, containing MgSiO$_3$ and Fe clouds, including scattering. The physics of this model is already included in \prt, with a few parameters such as the settling parameter f$_{sed}$, the vertical diffusion coefficient K$_{zz}$, cloud base pressures, and cloud mass fraction for each of the cloud species. We see that the given cloud model dampens the emission spectra of up to 30\% in between 2 and 5 \mum and for wavelengths higher than 10 \mum (clouds affect the spectra below 2 \mum a lot more). In the region between 5 and 10 \mum where a lot of molecular features are located for the MRS, the spectra seem less affected. When simulating observations of cloudy spectra with \msim, the clouds reduced the S/N of the cross-correlation molecule detection by up to 20\%. Molecules such as \nh3 and \ch4 are affected more since the clouds show a higher extinction at the wavelengths where these molecules have features. 

As is often the case, we can expect that dedicated data-driven algorithms will improve the sensitivity of the instrument to faint sources such as exoplanets (e.g. recent high contrast imaging results with HST by \citealt{zhou_pds70_hubble_2021}). For the MRS there are two avenues of the data reduction that should be explored. First, moving the analysis from the data cube to the detector space can benefit our science case by decreasing interpolation effects and correlated noise introduced by the cube reconstruction algorithm. Second, one of the crucial steps in direct imaging is modelling the stellar PSF and removing it from the data to improve the contrast to fainter companions \citep{Stolker_2019_pynpoint}. This is handled in a very simplistic way in our analysis, as we did not have the ability to simulate realistic behaviour of the instrument PSF. Once in orbit, we will have a better grasp on the effect the optical stability will have on the cross-correlation and can test PSF modelling ideas as for example in \cite{Ruffio2019} or \cite{ygouf2017_nirspec}.

By design, molecular mapping - especially in space - is quite robust to false positive detection since it relies on very specific patterns in the data that instrument systematic errors cannot replicate easily, contrary to some detector systematic trend in the continuum that could be interpreted as a broad molecule absorption feature. Additionally, compared to ground based observations, contamination due to earth's atmospheric absorption features will not be an issue for the MRS. With upcoming programmes from Guaranteed Time Observations \citep[GTO, ][]{GTO_TWA27} and Early Release Science \citep[ERS, ][]{ERS_exoplanets}, we will have the opportunity to test this technique on targets with relaxed contrast requirements and compare the results against state of the art Bayesian fitting methods.

\subsection{Cross-correlation and retrievals}

Detecting the existence of molecules in an atmosphere based on the cross-correlation signal is very useful to verify that the planet exhibits the expected trends that are predicted from exoplanet atmosphere theory. However, the end goal is to measure the abundance of a variety of molecules in the atmosphere. With such measurements one can build a deeper understanding of the physical properties of the atmosphere and compute elemental abundance ratios such as C/O and N/O that can both place exoplanets in context with each other and shed light on formation pathways \citep{Nowak_2020_BetaPicb}. 

Having transformed the cross-correlation into a likelihood function, as in \cite{brogi_retrieving_2019} we argue that the next logical step is performing a free retrieval within a Bayesian framework. Since it is expected that the high-pass filtering process removes much of the information regarding the temperature and gravity, we would place tight priors on these parameters and let the retrieval focus on estimating the abundances of molecules. This is justified from previous attempts of estimating the temperature and gravity with a grid of models for beta Pictoris b \citep{Hoeijmakers2018} and HIP65426 b \citep{Petrus2020}, that show that no clear optimum is reached and the cross-correlation tends towards high values for both parameters.

But we are not limited to just using the MIRI-MRS medium resolution spectra to constrain our models. All data, including near-IR spectra, flux measurements, and especially MIRI Coronagraphic measurements, that cover part of the MRS wavelength, can be included as additional likelihood terms \citep{brogi_retrieving_2019}. With better quality data from instruments such as GRAVITY \citep{2017_GRAVITY} and NIRSpec we can expect the retrievals to take advantage of the complementary information that different wavelength ranges and observing modes offer. For example,  \citealt{Burningham2021} demonstrated how both near and mid infrared data are required to constrain atmosphere and cloud properties.

From Fig.~\ref{fig:MIRISIM example} and Fig.~\ref{fig:Molecule features} we can observe that the features of molecules that are probed through molecular mapping are not necessarily coinciding with the broadband molecule features that are currently the focus of atmosphere retrievals. As an example in the mid-IR the \nh3 feature between 10 and 11 \mum is the focus of most observations and hence the MIRI Coronagraphs have filters at 10.65 and 11.4 \mum to be able to measure the \nh3 feature from the colour difference. For molecular mapping \nh3 has continuous and identifiable features from 5 to 10 \mum and the 10-11 \mum wavelength range where the broadband absorption feature is located does not show any particularly unique behaviour in the high pass filtered spectra. This is where the complementary information stems from. The retrieval routine can use information from both broadband features and the high frequency features simultaneously to constrain parameters and potentially overcome bias effects, such as using a limited part of the wavelength, that might affect the posterior distribution. The evaluation of the performance and gain in adding mid-IR medium resolution spectra to an atmospheric retrieval analysis will be explored in depth as part of future work.

\subsection{Sensitivity of cross-correlation to calibration errors}
The cross-correlation signal of a planet spectrum with a forward model enables the detection of the planet under the dominant stellar flux. Here, we evaluate the sensitivity of the cross-correlation signal to instrumental effects that could arise with JWST in orbit. 
While extended MRS calibration campaigns on the ground and during commissioning with JWST/MIRI in space will be able to correct many of the problems that are found in the data, there are uncertainties in the photometry and wavelength calibration that might not be possible to address. 
For example, fringing of the detectors creates subtle deviations from the calibration model due to the dependence of the fringes to the exact phase of the incoming beam. It has been verified that the uncertainty of these errors remains within acceptable limits, a process which will be repeated when the instrument is in orbit. 
Furthermore, in case one calibration step, such as fringing or spectro-photometric calibration, in the pipeline is very critical for the proposed science of this work, additional calibration effort might be required. There is already a plan in place to improve the instrument performance during Cycle 1 with dedicated calibration observations\footnote{\url{https://www.stsci.edu/jwst/observing-programs/calibration-programs}}. 

\begin{figure*}[h!]
    \centering
    \includegraphics[width=\linewidth]{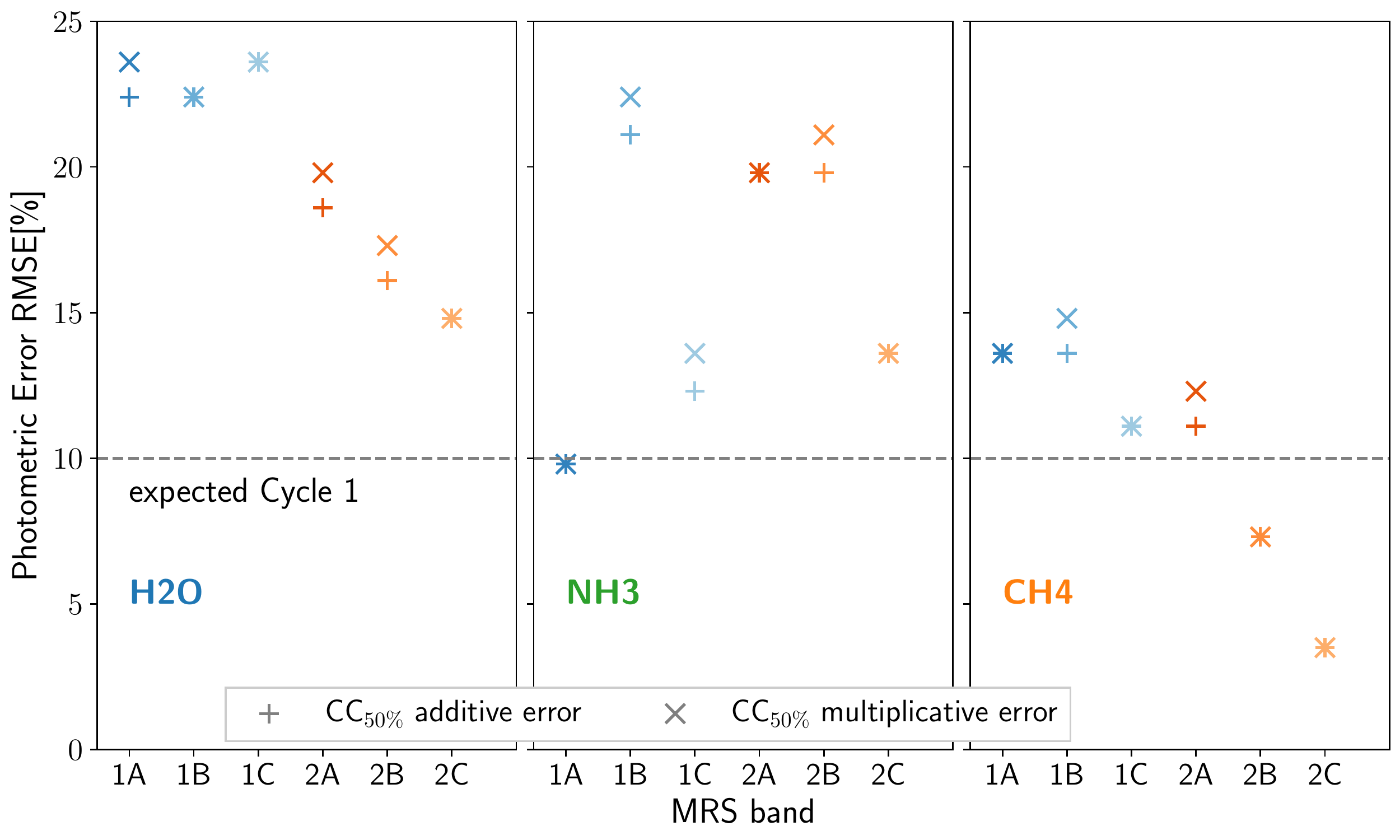} 
    
    \caption{Sensitivity of cross-correlation to  additive and multiplicative photometric errors of single molecule spectra: the plotted points represent the root mean square error with respect to the noiseless spectrum that is required such that the cross-correlation peak is reduced by 50\%. The dashed line indicates the expected photometric error in Cycle 1.
    }
    \label{fig:Sensitivity}
    \vspace{-1em}
\end{figure*}

We model both photometric and wavelength distortion errors with a simple approach, regarding individual, independent spatial frequencies that systematic errors might manifest in, and control the amplitude of the error as a function of the root mean square error (RMSE) of the residuals. 
We begin with an exoplanet spectrum model computed with \prt and convolve it to the mean resolution of each MRS sub-band and bin it to the detector sampling.
Then an additive or multiplicative photometric error is applied, or the wavelength coordinates of the spectrum with respect to the forward model is distorted.

Since the wavelength coordinate is recorded on the detector we consider each wavelength bin to represent a pixel on the detector, and spatial frequencies of detector pixel-wise dependent errors are simply transformed one-to-one onto the wavelength grid of our spectrum. 
This is not entirely correct since in the MRS detectors wavelength and spatial coordinates are curved on the detector but we consider this approach suitable for tackling this question. 
We then go through all the spatial frequencies that are constrained within the size of the detector, with the smallest frequency being a gradient over the whole array, and the highest frequency being a modulation with a period of one pixel. In Fig.~\ref{error_examples} examples of the different errors generated are shown.

We then proceed to high pass filter the simulated noisy spectra and cross-correlate them with the noiseless spectrum, measuring the peak value of the cross-correlation. In Fig.~\ref{fig:Sensitivity} we plot the effect of photometric errors on the cross-correlation signal of different molecules for the MRS sub-bands of Channel 1 and 2. For this test we took the single molecule templates seen in Fig.~\ref{fig:Molecule features} and added the additive and multiplicative photometric noise and calculated the average cross-correlation signal over all the frequencies for each level of noise. We set as a upper limit of photometric error a 50\% RMSE, compared to 10\% expected in Cycle 1, and plot the RMSE required to decrease the cross-correlation by half. We observe a dependence of the sensitivity for the different molecules as a function the MRS sub-bands, as it is expected since the features change with wavelength. We can conclude that from the tested molecules with the given P-T profile, \ch4 is the most sensitive to photometric errors. Regarding the wavelength calibration error, we find that uncertainty of the current solution \citep[ground solution $\sim$ 0.1 of resolution element, ][]{Labiano2021} suggests that it does not affect the cross-correlation signal, contributing to maximum a 1-2\% reduction. Uncertainty in the absolute wavelength solution again does not affect the result since it would manifest as an error in the radial velocity shift of the planet.

\subsection{Best targets might yet come}
The current limitation of the demographics of directly imaged planets is the small number of detected sub-stellar, planetary-mass companions and two of the published direct imaging surveys \citep[GPI, SHINE][]{Nielsen_2019_GPI, 2020_SHINE_Vigan} indicate detection limits as a function of mass and semi-major axis of a few times the mass of Jupiter at tens of AU. Recently, the Young Suns Exoplanet Survey \citep[YSES][]{Bohn_2020, Bohn2021} discovered new planets at wide orbits as part of a sample of young solar-type stars. Given the large separation from their host, these planets would be prime targets for characterisation with the MRS and more might be discovered by the YSES survey. \cite{Carter2021} investigated the detection limits expected with the JWST coronagraphs and found increased sensitivity to Jupiter and even sub-Jupiter mass, widely orbiting giant exoplanets that are missed due to the limited sensitivity of ground based instruments. If detected, follow up characterisation of these planets could be performed with the MRS, greatly contributing to exoplanet demographics and atmosphere characterisation of a diverse sample of objects.

\section{Conclusion}
The James Webb Space Telescope will open the window to mid-infrared characterisation of giant exoplanet atmospheres at an unprecedented sensitivity. We show that the Medium Resolution Spectrometer on board JWST/MIRI enables the detection of molecules from the emission of directly imaged exoplanets, by taking advantage of the moderate spectral and spatial resolution to disentangle the planet signal from stellar contamination. Constraints derived at these wavelengths can complement other spectroscopic observations in the near-infrared, as well as mid-infrared coronagraphic flux measurements to shed light on the atmosphere physics, chemistry, and formation of these class of planets. 

\begin{acknowledgements}

We thank the anonymous reviewer for their helpful comments that improved the manuscript quality. GC and SPQ thank the Swiss National Science Foundation for financial support under grant number 200021\_169131. IA thanks the European Space Agency (ESA) and the Belgian Federal Science Policy Office (BELSPO) for their support in the framework of the PRODEX Programme.

This research has made use of the NASA Astrophysics Data System and the \texttt{python} packages \texttt{numpy} \citep{harris2020array}, \texttt{scipy} \citep{2020SciPy-NMeth}, \texttt{matplotlib} \citep{Hunter:2007} and \texttt{astropy} \citep{astropy:2013, astropy:2018}.
\end{acknowledgements}

\bibliography{mrs_exoplanets.bib}
\bibliographystyle{aa.bst}

\begin{appendix}

\section{MRS visibility tool}\label{sec:visibility_tool}
We adapted the coronagraphic visibility tool\footnote{\url{https://jwst-docs.stsci.edu/jwst-other-tools/jwst-target-visibility-tools/jwst-coronagraphic-visibility-tool-help}} by rotating the coordinates to match the rotation of the MRS field with respect to the MIRI Imager. Additionally we enable plotting the different available dither patterns and allow for offsets of the target (usually the star). We plan to submit this tool to the developers of the coronagraphic visibility tool.

\begin{figure}[h!]
    \centering
    \includegraphics[width=\linewidth]{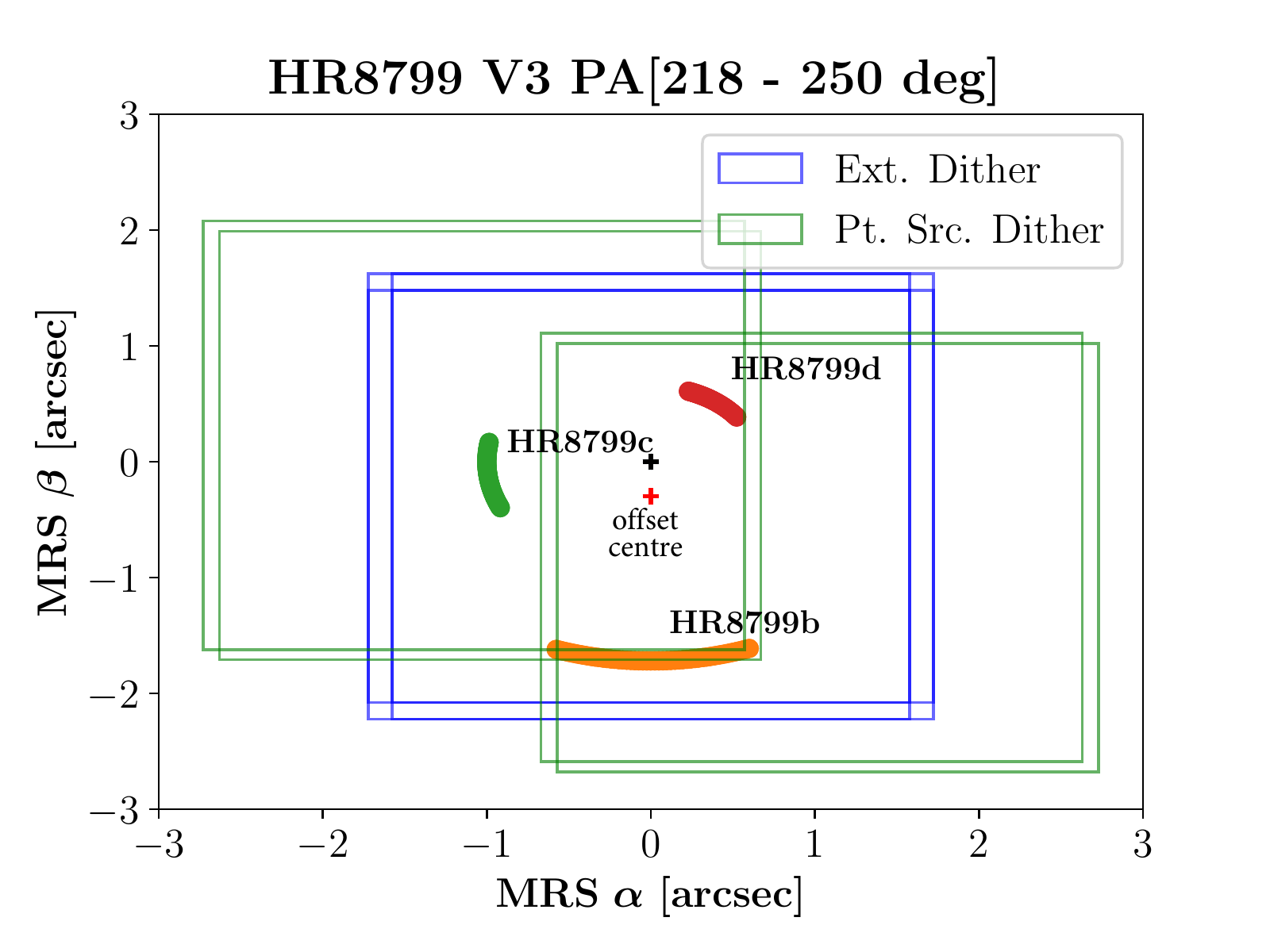}
    \caption{Example output of the MRS visibility tool. In order to properly sample the PSF while ensuring all targets in the HR 8799 system remain in view, we choose to use the CH1 extended source (Ext.) 4-point dither pattern, with a -0\farcs3 offset along the MRS $\beta$ axis. The nominal point source (Pt. Src.) 4-point dither pattern would provide better spatial sampling but leave one or more planets outside the FoV for half the observations.}
    \label{fig:dither visibility}
    \vspace{-1.5em}
\end{figure}

\section{Potential planetary-mass targets for molecular mapping with the MRS}\label{ap:targets}

\begin{table*}[h!]
    \centering
    \caption{Sub-stellar, planetary mass objects that would be observable by MIRI-MRS. The catalogue was retrieved from the Exoplanet.eu (\url{exoplanet.eu}) database and includes all bound objects regarded as planets (excluding verified Brown Dwarfs) with estimated mass up to 20 M$_{J}$. Planets that are not resolved from their host star by the MRS and planets orbiting at wide separations (>400 AU) are excluded.}
    \begin{tabular}{l|lllllll}
    \toprule
        \textbf{Target}& \textbf{T$_{\mathbf{eff}}$} & \textbf{a} & \textbf{Mass} & \textbf{Distance}& \textbf{Ang. Sep.} & \textbf{Reference} \\
                      &\textbf{[K]}            & \textbf{[AU]}& \textbf{[M$_{\mathbf{J}}$]}& \textbf{[pc]}& \textbf{[as]}& &\\[2pt]
        \hline \vspace{-0.85em}\\ 
        HR 8799 d                & 1200          & 27.0     & 8.3$\pm0.6$     & 41.3     & 0.65    & \cite{Wang_2018_HR8799}  \\
        HR 8799 c                & 1200          & 42.9     & 8.3$\pm0.6$     & 41.3     & 0.94    &  \cite{Wang_2020_HR8799c}  \\
        HR 8799 b                & 1100          & 68.2     & 7$_{-2}^{+34}$   & 41.3     & 1.72.  &  \cite{Barman_2015_HR8799b} \\
        GJ 504 b                & 550           & 43.5   & 4.0$_{-1}^{+4.5}$ & 17.5     & 2.48    &   \cite{bonnefoy_gj_2018}\\
        HD 95086 b             & 1050          & 61.7     & 2.6$\pm0.4$     & 90.4      & 0.62    &   \cite{deRosa2016} \\ 
        HIP 65426 b            & 1600          & 109.2    & 9$\pm3$         & 111.4     & 0.83    &  \cite{Cheetham_HIP654_2019}\\ 
        2M 0219 b         & 1700          & 160      & 13.5$\pm1.5$    & 39.4     & 3.96     &   \cite{Artigau2015_2M0219b} \\ 
        YSES-1 b$^{**}$        & 1700          & 162      & 14$\pm3$         & 94.6     & 1.71    & \cite{Bohn_2020}\\ 
        YSES-1 c$^{**}$        & 1200          & 320      & 6$\pm1$          & 94.6     & 3.4   & \cite{Bohn_2020}\\
        YSES-2 b        &  - & 115      & 6.3$_{-0.9}^{+1.6}$          & 110     & 1.05   & \cite{Bohn2021}\\
        2M 2236 b              & 1200          & 230      & 12.5$\pm1.5$    & 63      & 3.7     &\cite{Bowler_2236b_2017}\\
        ROXs 42B(AB) b  & 1900 &  140 &   9$\pm3$ & 135 & 1.2 & \cite{Daemgen2017} \\
        HD 203030 b & 1050 & 487 & 11$_{-3}^{+4}$ &  41&  11.9 & \cite{Miles_P_ez_2017}\\
        2M 0103(AB) b & - & 84 & 13 $\pm 1$ &  47.2 & 1.7 & \cite{Delorme2013} \\
        AB Pic b & 2000 & 275 & 13.5 $\pm 0.5$  &  47.3 & 5.5 & \cite{Bonnefoy2010_ABPicb} \\
        1RXS 1609 b &1800 & 330 & 14$_{-3}^{+2}$ & 145 & 2.2 & \cite{Lafreniere2010}\\
        CT Cha b$^{**}$ & 2500 & 440 & 17 $\pm6$ & 165 & 2.6 & \cite{Wu2015_CTChaB}\\
        GSC 6214-210 b & 1700& 320& 17 $\pm 3$ & 145 & 2.2 & \cite{Pearce2019}\\
        51 Eri b$^*$ &  750 & 11.1 & 2.9$\pm0.3$& 29.4 &  0.5 & \cite{Samland_2017}\\
        $\kappa$ And. b$^*$         & 1850          & 100      & 13$_{-2}^{+12}$ & 50      & 1.06   &   \cite{Wilcomb2020} \\ 
        \midrule
        & \multicolumn{6}{c}{\textbf{GTO/ERS Targets}}\\
        \midrule
        $\beta$ Pic b     & 1750          & 9.2        & 9.8$\pm2.7$  & 19.3      & 0.5     & \cite{Nowak_2020_BetaPicb} \\
        2M 1207 b              & 1600    & 42          & 4$_{-1}^{+6}$& 52.4      & 0.78    &  \cite{Patience2010} \\
        VHS 1256 b        & 1000          & 102         & 19$\pm5$     & 22.2      & 8.1     &  \cite{2020_Zhou_vhs1256b}\\
        Ross 458 c             & 650      & 1160        & 11.3$\pm4.5$ & 63.0      & 102     & \cite{Burgasser2010}\\
        
        \bottomrule
    \end{tabular}
    
    \label{tab:targets}
    \tablefoot{\tablefoottext{*}{The host stars of these sub-stellar companions are over the bright source limit of the MRS detectors and can not be observed until partial saturation of the detectors is allowed.}, \tablefoottext{**}{JWST GO Cycle 1 accepted proposal to be observed with the MRS or LRS.}}
\end{table*}

\section{Log-likelihood ratio tables}

\begin{table*}[h]
    \centering
\caption{Log-Likelihood Ratio test statistic values for testing the existence of individual molecules in the atmosphere of HR8799 b. Numbers in bold indicate a significant value for a 99.9\% (3 $\sigma$) confidence level (critical value > 10.384).
\label{lLR_hr8799b}}
\begin{tabular}{lrrrrrrrrrr}
\toprule
{} &     1A &     1B &     1C &     2A &     2B &     2C &     3A &     3B &    3C &  Abundance \\
Molecule &        &        &        &        &        &        &        &        &       &            \\
\midrule
H2O      &  \textbf{920.6} & -192.8 &  -59.4 &  -16.3 &  \textbf{323.2} &   \textbf{22.2} &  \textbf{146.5} &  \textbf{366.3} &  \textbf{39.3} &       -2.1 \\
CO       &  \textbf{184.4} &   -0.0 &    0.0 &   -0.0 &   -0.0 &   -0.0 &   -0.0 &   -0.0 &   0.0 &       -2.8 \\
CO2      &   -0.2 &    0.0 &    0.0 &   -0.0 &    0.9 &   -0.1 &    2.5 &   \textbf{36.5} & -21.0 &       -5.3 \\
CH4      &  \textbf{314.1} &  \textbf{289.9} &  \textbf{585.5} &  \textbf{319.3} &  \textbf{335.7} &   \textbf{57.5} &    3.3 &    0.4 &  -1.9 &       -2.4 \\
NH3      &   \textbf{12.3} &   \textbf{22.7} &  -38.3 &   -9.9 &    0.4 &  \textbf{139.4} &   \textbf{72.4} &   \textbf{58.2} &   4.3 &       -5.0 \\
H2S      &   -0.0 &    0.6 &  -10.7 &    9.7 &    2.3 &   -0.0 &   -0.0 &   -0.0 &  -1.6 &       -3.3 \\
PH3      &   -7.5 &    0.0 &    0.3 &   \textbf{10.4} &   \textbf{44.3} &   \textbf{61.1} &   -0.2 &   -0.5 &  -0.0 &       -5.0 \\
\bottomrule
\end{tabular}

\end{table*}

\begin{table*}[h]
    \centering
\caption{Log-Likelihood Ratio test statistic values for testing the existence of individual molecules in the atmosphere of GJ504 b. Numbers in bold indicate a significant value for a 99.9\% (3 $\sigma$) confidence level (critical value > 10.384). 
\label{lLR_gj504b}}
\begin{tabular}{lrrrrrrr}
\toprule
{} &      1A &     1B &     1C &     2A &     2B &     2C &  Abundance \\
Molecule &         &        &        &        &        &        &            \\
\midrule
H2O      &  \textbf{1752.2} &  \textbf{932.7} &  \textbf{321.4} &   \textbf{84.4} &   \textbf{37.7} &  -14.9 &       -2.0 \\
CO       &     0.0 &    0.0 &    0.0 &    0.0 &    0.0 &    0.0 &       -10.0 \\
CO2      &     0.0 &    0.0 &    0.0 &    0.0 &    0.0 &    0.0 &       -10.0 \\
CH4      &   \textbf{152.0} &  -65.3 &  \textbf{139.6} &  \textbf{465.5} & -252.5 &    2.3 &       -2.3 \\
NH3      &   \textbf{245.5} &  \textbf{681.0} &  \textbf{137.0} &  \textbf{339.1} &  \textbf{438.2} &  \textbf{652.6} &       -2.9 \\
H2S      &     0.0 &   -0.0 &   \textbf{15.6} &   -4.5 &   -7.9 &   -0.2 &       -3.3 \\
PH3      &     0.0 &    0.0 &    0.0 &    0.0 &    0.0 &    0.0 &       -8.0 \\
\bottomrule
\end{tabular}
\end{table*}

\section{Error modelling}
All errors considered here are modelled as sine function of a given frequency $\nu$ and amplitude $A$ 
$$\epsilon(A, \nu) = Asin(2\pi \nu \lambda)$$
The RMSE for additive errors is related to the amplitude as:
$$RMSE = \frac{|A|}{\sqrt{2}}, $$

and is scaled to the maximum of the signal in each sub-band. The maximum is chosen since the data are normalised to the maximum after the high-pass filtering.

Additive photometric errors are considered as modulation in the measured DN/s due to effects independent of the flux on the detector, such as detector odd-even row effect \citep{Rieke2015b}. These errors can simply be added to the spectrum as
$$f_{additive} = f_{planet} + \epsilon(A, \nu).$$

Multiplicative photometric errors on the other hand have an effect on the transmission of the signal and scale with it. Such effects are similar to detector fringing, detector gain and spectro-photometric conversion uncertainty. These errors are multiplied to the spectrum as a transmission function 
$$f_{multiplicative} = f_{planet} (1 + \epsilon(A, \nu)).$$

For these errors, the amplitude was optimised such that the measured RMSE of the residuals was equal to the chosen value. Finally, wavelength distortion errors are again additive systematic errors and they perturb the wavelength coordinates on which the spectrum was sampled. In order to implement this, we took the nominal wavelength grid of each sub-band and added a systematic error creating a new wavelength grid. The high resolution spectrum was then sampled on the new wavelength grid:

$$\lambda' = \lambda + \epsilon(A, \nu).$$

\begin{figure}[H]
    \centering
    \includegraphics[width=\linewidth]{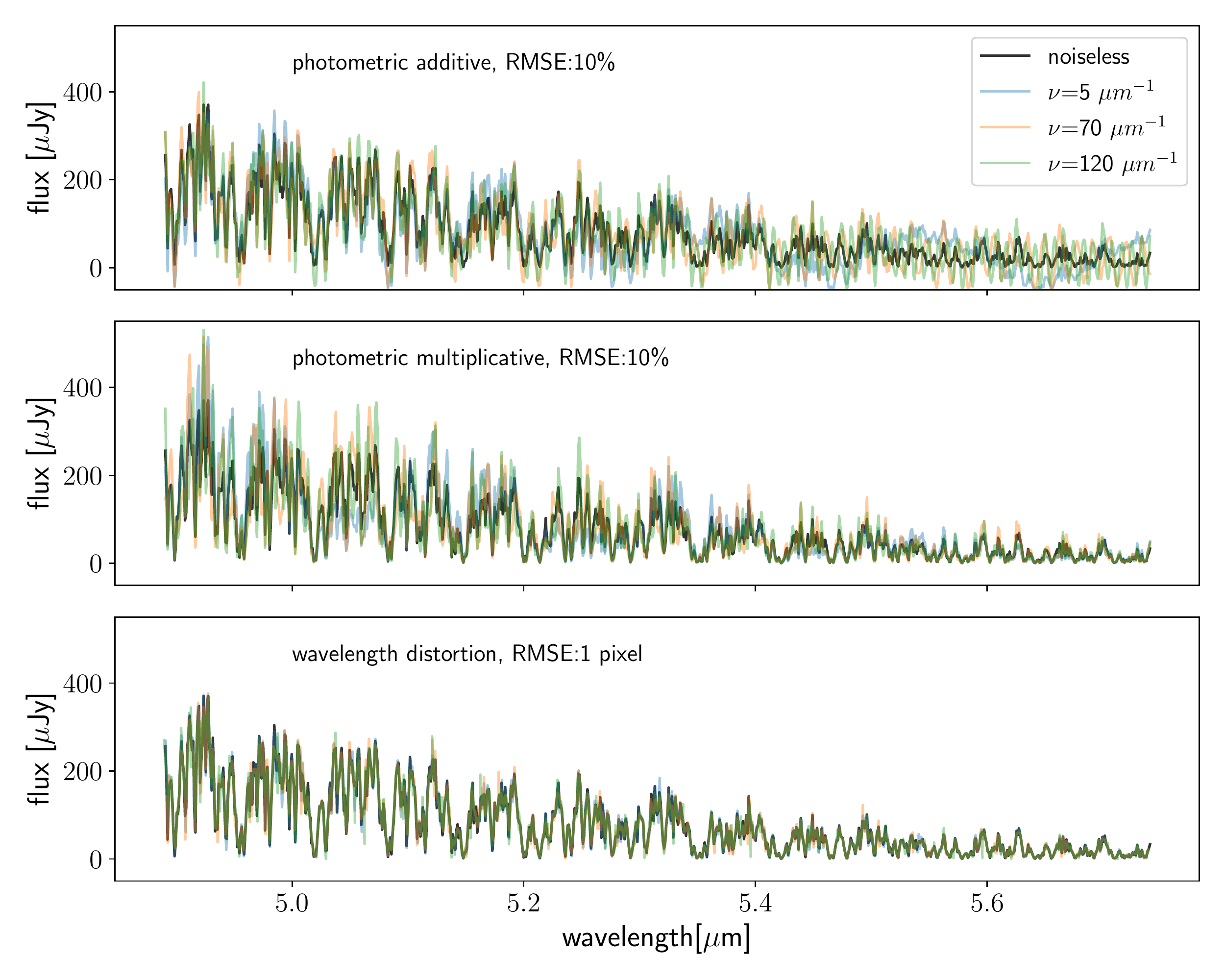} 
    \caption{Example of errors added to the noiseless spectrum with additive (top) multiplicative (middle) photometric errors and wavelength distortion (bottom).}
    \label{error_examples}
    \vspace{-1em}
\end{figure}

\end{appendix}

\end{document}